\documentclass[a4paper,12pt]{article}

\usepackage{jheppub} 

\usepackage[T1]{fontenc} 
\usepackage{bm,latexsym,amsmath,amssymb,amsfonts,mathtools,mathrsfs}
\usepackage{comment}
\usepackage{color}

\usepackage{bm}
\usepackage{hyperref}
\usepackage{framed}
\usepackage{subfigure}
\usepackage{hyperref}   
\usepackage{graphicx}
\usepackage{makecell}

\expandafter\let\csname equation*\endcsname\relax 
\expandafter\let\csname endequation*\endcsname\relax 
\usepackage{amsmath}
\usepackage{color}

\usepackage{hyperref}
\hypersetup{
    colorlinks=true,
    linkcolor=blue,
    filecolor=magenta,
    urlcolor=blue,
}
\usepackage{float}
\usepackage{ifthen}
\usepackage{xspace}
\usepackage{relsize}
\usepackage{yhmath}
\usepackage[left]{lineno}

\newcommand{\digon}{\mathbin{\raisebox{0.12em}{$\boldsymbol\frown$} \hspace{-1.16em} \raisebox{-0.12em}{$\boldsymbol\smile$}}}

\begin{document}

\title{On the global Gaussian bending measure and its applications in stationary spacetimes} 
\author[a,1]{Zhen Zhang\note{Corresponding authors.}}
\author[b,c,1]{, Rui Zhang}

\affiliation[a]{Key Laboratory of Particle Astrophysics, Institute of High Energy Physics, Chinese Academy of Sciences, 
19B Yuquan Road, Beijing 100049, People’s Republic of China}
\affiliation[b]{Theoretical Physics Division, Institute of High Energy Physics, Chinese Academy of Sciences, 
19B Yuquan Road, Beijing 100049, People’s Republic of China}
\affiliation[c]{Shanghai Key Laboratory of Deep Space Exploration Technology, Shanghai Institute of Satellite Engineering, No. 3666 Yuanjiang Road, Shanghai 201109, People’s Republic of China}

\emailAdd{zhangzhen@ihep.ac.cn}
\emailAdd{rui.z@pku.edu.cn}

\date{\today}

\abstract{
Modified gravity theories have been suggested to address the limitations of general relativity, each exhibiting differences, particularly in their strong-field limits. Nonetheless, there lacks effective means to distinguish or test these theories through local strong-field measurements. In this work, we define a global Gaussian bending measure over singular spacetime regions, establish a corresponding global theory, and demonstrate its applications in a general stationary spacetime. The global theory is based on differential geometry, rather than on specific gravity theories, allowing it to depict various physics within general relativity and beyond. For example, it can be applied to describe the gravitational bending of massless or massive messengers, such as photons, neutrinos, cosmic rays, and possibly massive gravitational waves predicted in certain theories of gravity. Besides, the global theory is applicable to any stationary spacetime regions outside a rotating black hole. As an instance of its direct applications, we investigate the highly-curved spacetime effects of the black hole in its immediate surrounding regions and design local strong-field experiments involving different shapes of singular lensing patches. New means can be therefore anticipated to be developed according to the global theory to differentiate between different gravity theories and test them in their strong-field regions.
}

\maketitle


\section{Introduction}

Nowadays, Einstein's theory of general relativity (GR) is still the most successful theory of gravity.
Since its birth~\cite{Einstein1916}, it has made numerous profound theoretical predictions~\cite{Weinberg:1972kfs,Chandrasekhar1983,Rindler2006book,Carroll2014}.
The first observation of the gravitational bending of light
goes back to 1919 when Sir Eddington's expedition gave the
striking verification of Einstein's prediction during a Solar Eclipse~\cite{Eddington1920},
which is vital to the establishment of GR. Since then, there has been substantial progress on the mathematical formulation concerning the bending of light~\cite{Bohr2015,Darwin59,VE2000,RI2007,Ishak:2008ex,Arakida:2011ty,Gibbons2008,Werner2012,Ishihara2016,Ishihara2017,Arakida:2017hrm,Takizawa2020,Sanchez:2023ckq,Ovgun:2018fnk,Ovgun:2019wej,Zhang:2021ygh,Qiao2022,Hioki:2009na}.
Apart from photons, massive messengers such as neutrinos, cosmic rays, and possibly massive gravitational waves
may also play a crucial role in exploring the universe through the effects of gravitational bending.
So far, many modified gravity (MG) theories have been suggested to overcome certain perceived limitations of GR~\cite{Lovelock1971,KYY2011,PK2013,Glavan20,SZ2021,MA2022,QG2023,Hohmann:2022mlc,Shao2023}, 
such as the presence of singularities~\cite{Weinberg:1972kfs,Chandrasekhar1983,Rindler2006book,Carroll2014}, and to incorporate quantum gravitational effects~\cite{Burgess2004,Donoghue:2015hwa}. 
Some of these MG theories can explain astronomical observations, such as the phenomena related to gravitational bending~\cite{Pantig:2024kfn,Atamurotov:2022knb,Chen:2022ynz,Liu:2022hbp,Guo:2022muy,Fu:2022yrs,Qiao:2022nic,Perivolaropoulos:2023zzc,Mustafa:2022xod,Pantig:2022whj,Qiao:2021trw,Ali:2022ubd,Adler:2022pqw,Pantig:2022toh,Lake:2021gbu}.
Here, some of these observations involve `strong-field experiments'.
However, in these experiments, all measurements are performed by the observers in weak-field or almost flat regions of spacetime, similar to the asymptotically flat regions far from a black hole. Although these gravity theories may exhibit significant differences in the strong-field limit~\cite{PK2013,Glavan20,SZ2021,MA2022,QG2023,Bernardo:2022acn}, the effects of those differences could be lost during the weak-field measurements.
Hence, it is important to test various theories of gravity through local strong-field experiments, requiring local observers (or detectors) to make measurements in highly-curved regions.

Recently, a (local) theory has been proposed based on the Gaussian bending (or deflection) angle, 
without relying on particular shapes of lensing patches or specific models of spacetime~\cite{Zhang:2021ygh}.
The Gaussian bending angle is actually an extension of the usual (or traditional) deflection angle~\cite{Weinberg:1972kfs,Chandrasekhar1983,Rindler2006book,Carroll2014}.
Note, the usual deflection angle is defined as the Euclidean intersection angle between the incident and outgoing light rays at spatial infinities where the spacetime become flat \cite{Lake:2007dx}, which is actually a coordinate angle dependent on the coordinate system chosen, and in highly curved regions, this angle will lose its measurability in the strong-field measurements by the local observers for the corresponding light rays at the incident and outgoing spacetime points in these regions~\cite{RI2007,Ishak:2008ex,Arakida:2011ty}.
Alternatively, Gibbons and Werner proposed another method to define the asymptotic deflection angle of light, known as the weak deflection angle, which assumes the receiver and source are in an asymptotically flat region \cite{Gibbons2008,Werner2012}. 
Thereafter, their idea has been extended to explore the effects of finite distance on light bending, where both the receiver and source are within a finite distance from the involved gravitational lens \cite{Ishihara2016,Ishihara2017}.
Other similar measures have also been defined using geodesic polygons in \cite{Arakida:2017hrm,Takizawa2020,Sanchez:2023ckq}.
For example, the measure in \cite{Arakida:2017hrm} is specified for a particular tetragon whereas the measure in \cite{Sanchez:2023ckq} for a specific triangle.
Currently, various bending measures and their applications are actively being explored within various theories of gravity \cite{Sanchez:2025dvb,Takizawa:2020dja,Jusufi:2017hed}.
However, none of these measures is applicable to singular polygons.
Additionally, the local Gaussian bending measure is not reliant on any specific polygons.

In the local theory of Gaussian bending, a geometrisation is globally performed on the propagation of massless or massive messengers, 
and the definition of the gravitational bending angle of these messengers is generalised to any static curved spacetime regions. 
In a given oriented surface $\Sigma$, $D\subset\Sigma$ is supposed to be physically simple, connected region
whose boundary $\partial{D}$ is a closed regular curve with induced orientation from ${D}$; see Figure~\ref{fig:gdp1} for more details, especially in the region $D$ without singularities. 
The boundary curve $\partial{D}$ should be composed of finitely many piecewise smooth simple geodesic segments~\cite{Zhang:2021ygh} without self-intersections~\cite{Carmo16,Chern00}.
It can be parametrised in the right-handed direction by the arc length $\lambda$.
Then, set $\bar{\lambda}=l-\lambda$, where $l$ is the total arc length of the closed boundary $\partial{D}$. 
For an observer at the point $\lambda=\lambda_{\rm 0}$, two vectors can be obtained by the parallel transport of any vector at the point $\lambda=0$ of some source along two geodesics, respectively.
The angle between these two transported vectors, as measured by this observer, is defined as the {\it Gaussian bending angle}, 
i.e., $\alpha_{M}=\bar{\varphi}\left(\bar{\lambda}_{0}\right)-\varphi\left(\lambda_{0}\right)$. In an asymptotically flat spacetime, it reduces to the usual deflection angle when both the observer and source are located in the distant flat regions of the spacetime~\cite{Zhang:2021ygh}, where $\bar{\varphi}\left(\bar{\lambda}\right)$ and $\varphi\left(\lambda\right)$ denote the two angles from a given axis to the two transported vectors at $(\lambda,\bar{\lambda})$, respectively. 
Moreover, this angle can be further expressed as~\cite{Zhang:2021ygh}
\begin{eqnarray}
\label{eq:GaussianCurvature}
\begin{array}{rcl}
\displaystyle
\alpha_{M}=\bar{\varphi}\left(\bar{\lambda}\right)-\varphi\left(\lambda\right)=-\int\!\!\!\int_{D}\,K\mathrm{d} \sigma,
\hspace*{0mm}
\end{array}
\end{eqnarray}
where $\mathrm{d} \sigma$ represents the area element, $K$ is the Gaussian curvature.
Here, the region ${ D}$ is known as a \textit{lensing patch}, on which the measurement can be performed by local observers. 
This bending formula~\eqref{eq:GaussianCurvature} generalises that for the weak deflection angle~\cite{Gibbons2008,Werner2012},
which can be confirmed through a comparison with a simple and straightforward expression for the latter presented in~\cite{Ovgun:2018fnk, Ovgun:2019wej}.
Crucially, the bending formula~\eqref{eq:GaussianCurvature} can be utilised to depict the propagation of massless and massive messengers in strong gravitational fields.

The local theory and its applications are partially founded on the Gauss-Bonnet theorem~\cite{Zhang:2021ygh}. 
However, the Gauss-Bonnet theorem is only applicable if the region ${D}$ contains no curvature singularities{\footnote{In differential geometry, mathematicians have systematically studied a few types of singular points, including the conical singularities as well as those found at end points or branch points, and they extend the Gauss-Bonnet formula to account for the contributions arising from regions containing these points~\cite{Buzano_2019,CohnVossen1935KrzesteWU,Huber1958,Hartman1964GEODESICPC,Finn1965,JORGE1983203,Li1991CompleteSW,Troyanov1991PrescribingCO,Chen1995WhatKO,Shiohama2010TOTALCA,Nguyen2012}. In theories of gravity, singularities refer to the ``curvature singularities” at which the Kretschmann scalar become infinite. Currently, there is no extended Gauss-Bonnet formula that is applicable to regions with ``curvature singularities”, even when dealing with a two-dimensional differentiable manifold (or surface). }}~\cite{Carmo16,Chern00}. 
Generally, it is inevitable to investigate the stationary regions featuring curvature singularities when dealing with a gravitational system~\cite{Penrose1964,Hawking1970} like a rotating black hole in GR. 
In this work, we are endeavoring to define a global\footnote{The word ``global'' has the same meaning as that in the term ``global Gaussian-Bonnet theorem'', distinguishing it from other interpretations.} Gaussian bending measure and establish its global theory. 
First of all, we introduce the global version of Gaussian bending in the spacetime regions with curvature singularities. Then, we illustrate how to apply the global theory in a general stationary spacetime and demonstrate the strong-field effects of a rotating black hole on the Gaussian bending of light. Subsequently, we discuss the prospects for the practical applications of the global Gaussian bending measure in the exploration of the universe. 
Finally, we summarize our findings.

\section{A generalised bending measure and its global theory} 
\label{sec:2} 

\subsection{Gaussian bending measure: global definition and bending formula}
\label{sec:2.1} 

Singularities are inevitable in GR~\cite{Penrose1964,Hawking1970}, although they may be hidden behind event horizons. 
When studying or probing the spacetime structure of a black hole in particular, we need to take into account the regions with singularities.
Especially in the local theory~\eqref{eq:GaussianCurvature}, if the lensing patch ${D}\subset\Sigma$ contains singularities, where $\Sigma$ is a physical surface on which a source, a black hole acting as a gravitational lens, and an observer are located, the Gaussian bending angle~\eqref{eq:GaussianCurvature} cannot be derived directly from the Gauss-Bonnet theorem, and hence also the weak deflection angle~\cite{Gibbons2008}, as the Gauss-Bonnet theorem no longer holds true in this case.

\begin{figure}
\centering
\includegraphics[width=1.0\columnwidth]{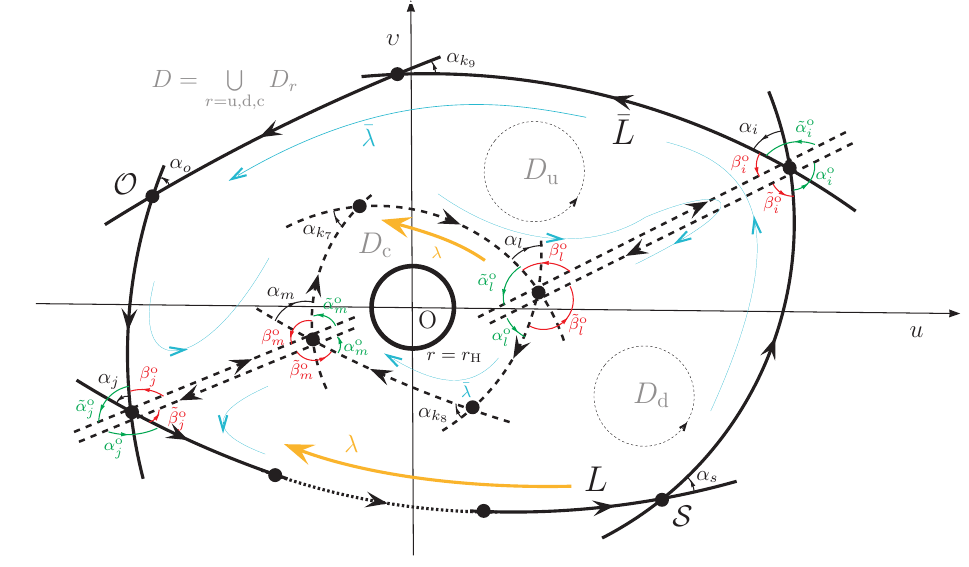}
\caption{
Illustration of global Gaussian bending over the lensing patch $D\subset\Sigma$, featuring a singularity located at the origin ${\rm O}$ on the physical surface $\Sigma$. The sub-region ${ D_{\rm c}}\subset{D}$ containing the singularity can be removed along a closed curve, denoted as $\partial{D}_{\rm c}$. The region that remains, $\mathring{D}$, is free of singularities, but it has a geometric ``hole'', with its boundaries $\partial{\mathring D}$ made up of geodesic line segments. As illustrated in the figure, four vertices (or points), $\gamma_{i},  \gamma_{j}, \gamma_{l},~{\rm and}~ \gamma_{m}$, can be chosen in such a way that the region $\mathring{D}$ can be cut into two parts, namely ${ D_{\rm u}}$ and ${ D_{\rm d}}$, along the line segments $C_{il}=\overline{\gamma_{i}\gamma_{l}}$ and $C_{jm}=\overline{\gamma_{j}\gamma_{m}}$. Here, $D=D_{\rm u}\bigcup D_{\rm d}\bigcup D_{\rm c}$. Note that the boundaries, ${\partial D_{\rm u}}$ and ${\partial D_{\rm d}}$, of the two parts can be jointly described by a parametrisation $\gamma:\,[0,l]\to{\partial D_{\rm u}}+{\partial D_{\rm d}}~(\supsetneq\partial{\mathring D})$, forming a closed curve, where $l$ is the total arc length of the closed curve $\gamma$. Assume that $\gamma$ is parametrised by arc length $\lambda$ in the right-handed direction, as marked by the brown arrows, whereas it is parametrised by arc length $\bar{\lambda}=l-\lambda$ in the left-handed direction, as indicated by the cyan arrows.  Let $\gamma\left(\lambda_{k}\right)=\gamma_{k}, {\rm for}~k=0, ... , \aleph$, be the vertices of $\gamma$, with $\lambda_{k}$ denoting the value of $\lambda$ at the $k$-th vertex. Suppose that the source and observer are located at points ${\cal S}=\gamma(\lambda=0)=\gamma(\lambda=l)$ and ${\cal O}=\gamma\left(\lambda=\lambda_{0}\right)$, respectively. All the symbols are detailed in Table~\ref{tab:symbol}. For a black hole, the singularity is hidden behind the event horizon located at a radius of $r=r_{\rm H}$, and thus, $\partial{D}_{\rm c}$ can be chosen as the intrinsic boundary at the event horizon. If $D$ is singularity-free, $D_{\rm c}$ is left empty, and $\partial{\mathring D}=\partial{D}$ is a simple closed curve.
}
\label{fig:gdp1}
\end{figure}

Let us consider the following situation. As depicted in Figure~\ref{fig:gdp1}, two light rays originate from a source at the point ${\cal S}~[=\gamma(0)=\gamma(l)]$ and travel along two curves, $L$ and $\bar{L}$, located on opposite sides of the singularity at the origin $\rm{O}=(0,0)$, before reaching the observer at the point ${\cal O}~[=\gamma\left(\lambda_{0}\right)]$, respectively. Here, $L$ and $\bar{L}$ consist of piecewise geodesic segments. 
A region, denoted as ${D}$, is bounded by $L$ and $\bar{L}$, and it contains the singularity at $\rm{O}$, forming a lensing patch.
This situation can be observed in gravitational lensing, 
where both $L$ and $\bar{L}$ represent light trajectories, or physically simple null geodesics without self-intersections.
In fact, as shown by Figure 11.9(b) in~\cite{Rindler2006book} or by the geodesic digon in our Figure~\ref{fig:digon}, 
there can be two possible trajectories of this kind for the photons from a light source to us on a physical surface $\Sigma$. 
Between the two light trajectories, there are two intersection points, namely the vertices ${\cal S}$ and ${\cal O}$.
We denote by $\partial{D}$ the boundary of the lensing patch $D$.
The external (interior) angles of $\partial{D}$ are actually the intersection angles between $L$ and $\bar{L}$ at points ${\cal S}$ and ${\cal O}$, 
denoted as $\alpha_{s}$ ($\beta_{s}$) and $\alpha_{o}$ ($\beta_{o}$), respectively.
In principle, each of these angles is measurable. 
However, we are unable to directly establish their relationships with the Gaussian bending angle~\eqref{eq:GaussianCurvature} through the application of the Gauss-Bonnet theorem,
due to the presence of a singularity in the lensing patch. 

Now, let us place a black hole at the origin $\rm{O}$, as illustrated in Figure~\ref{fig:gdp1}. 
Thus, there is a singularity at the origin in the lensing patch ${D}$.
Then, choose ${ D_{\rm c}}\subset{D}$ to be a geodesic polygon, meaning a polygon with its boundary $\partial{D}_{\rm c}$ composed of arcs of geodesics, and ensure that it contains the singularity.
For any vertex $\gamma_{p}$ of $\partial{D}$, there always exists a smooth line segment connecting this vertex to a corresponding point $\gamma_{q}$ on $\partial{D_{\rm c}}$. 
The corresponding point can be a vertex of $\partial{D_{\rm c}}$. If not, we can still treat it as a vertex with an external angle of zero.
Let $C_{pq}$ be the smooth line segment joining any two vertices $\gamma_{p}$ and $\gamma_{q}$, directed from the former to the latter, 
which may be non-geodesic. As illustrated in Figure~\ref{fig:gdp1}, we choose four vertices, i.e., $\gamma_{i}$, $\gamma_{j}$, $\gamma_{l}$, and $\gamma_{m}$.
By cutting along the boundary $\partial{D}_{\rm c}$ of the sub-region ${ D_{\rm c}}$ as well as the segments $C_{il}$ and $C_{jm}$, the lensing patch ${D}$ can be divided into three different regions: ${D_{\rm u}}$, ${D_{\rm d}}$, and ${D_{\rm c}}$, as shown in Figure~\ref{fig:gdp1}. Then, we can define a spacetime region with a geometric ``hole''\footnote{From here on, the concept of geometric ``holes'' is purely mathematical, resulting from the removal of singular regions like $D_{\rm c}$ from the lensing patch $D$.}, denoted as
\begin{eqnarray}
\label{eq:mathringD1}
\begin{array}{rcl}
\displaystyle
\nonumber
\mathring{D}=D-D_{\rm c}=\bigcup\limits_{r=\rm{ u,d}}D_r,
\end{array}
\end{eqnarray}
which contains no singularity. 
Let $\partial{D}_{r}$ represent the boundary of $D_r$ for $r={\rm u, d}$, respectively.
As Figure~\ref{fig:gdp1} shows, these boundaries can be parametrised in a unified manner by the arc length parameter $\bar{\lambda}$ 
along their tracks in the direction indicated by the cyan arrows, or by the arc length parameter $\lambda$ in the opposite direction, for example.
Then, let us define a {\it global Gaussian bending measure} as 
\begin{eqnarray}
\label{eq:edefinition}
\begin{array}{rcl}
\displaystyle
\mathring{\alpha}_{M}=\bar{\varphi}\left(\bar{\lambda}_{0}\right)-\varphi\left(\lambda_{0}\right),
\hspace*{0mm}
\end{array}
\end{eqnarray}
which is in analogy with the Gaussian bending angle $\alpha_{M}$. 
According to the existence and uniqueness of the parallel transport, one always has $\bar{\varphi}\left(\bar{\lambda}\right)-\varphi\left(\lambda\right)=\bar{\varphi}\left(\bar{\lambda}_{0}\right)-\varphi\left(\lambda_{0}\right)=constant$ along the uniformly parametrised curve~\cite{Zhang:2021ygh}, as required by the theorem of existence and uniqueness of differential equations~\cite{Carmo16,Chern00}. Meanwhile, one also has~\cite{Zhang:2021ygh,Carmo16}
\begin{eqnarray}
\label{eq:GBtoPT}
\begin{array}{rcl}
\displaystyle
\nonumber
\varphi\left(\lambda\right)-\bar{\varphi}\left(\bar{\lambda}\right)=\varphi\left(l\right)-\varphi\left(0\right)=\int\!\!\!\int_{\mathring{D}}\,K\,\mathrm{d} \sigma, 
\hspace*{0mm}
\end{array}
\end{eqnarray}
As a result, the {\it Gaussian bending formula}~\eqref{eq:GaussianCurvature} can be extended to become
\begin{eqnarray}
\label{eq:eGaussianCurvature}
\begin{array}{rcl}
\displaystyle
\mathring{\alpha}_{M}=\bar{\varphi}\left(\bar{\lambda}\right)-\varphi\left(\lambda\right)=-\int\!\!\!\int_{\mathring{D}}\,K\,\mathrm{d} \sigma,
\hspace*{0mm}
\end{array}
\end{eqnarray}
which is integrated over $\mathring{D}$ instead of $D$.
Note that the boundary $\partial{D}_{r}$ could be the intrinsic boundary of the physical surface $\Sigma$, such as the one at the event horizon of a black hole. 
Based on this bending formula, we are able to connect the global Gaussian bending measure $\mathring{\alpha}_{M}$ with the surface integral of the Gaussian curvature over the region $\mathring{D}$ with a geometric ``hole''. 
If the lensing patch $D$ is free of singularities, we define $\mathring{D}$ to be equal to $D$.
Subsequently, the global Gaussian bending measure $\mathring{\alpha}_{M}$~\eqref{eq:eGaussianCurvature} reduces to the Gaussian bending angle $\alpha_{M}$~\eqref{eq:GaussianCurvature}; the former is a generalisation of the latter.
Actually, the global measure can be further extended to the region with multiple geometric ``holes'', and hence the bending formula, as we will show soon. 
Besides, the Gaussian bending angle $\alpha_{M}$ is an extension of the usual deflection angle~\cite{Zhang:2021ygh}, and consequently, the generalised bending measure $\mathring{\alpha}_{M}$ as well. Hereafter, the Gaussian bending angle $\alpha_{M}$ will be termed as the local Gaussian bending measure;
if no specific emphasis is given, both the local and global measures could be referred to as the Gaussian bending measure.
Clearly, the bending formula~\eqref{eq:eGaussianCurvature} deeply reveals the equivalence between the bending measure and the surface integral of the Gaussian curvature over the chosen lensing patch. By definition, the Gaussian curvature $K$ is completely determined by properties intrinsic to spacetime, like spin $a$ and mass $M$ in a Kerr spacetime~\cite{Zhang:2021ygh}. Therefore, the Gaussian bending measure can serve as a potentially interesting probe into the intrinsic properties of spacetime.

General speaking, we can always define a global Gaussian bending measure $\mathring{\alpha}_{M}$ over the lensing patch $D$ with multiple singularities; more details on global Gaussian bending in singular spacetime regions are available in Appendix~\ref{app:T}. Once the global Gaussian measure $\mathring{\alpha}_{M}$ is determined, we can measure the spacetime effects on the gravitational bending of messengers by the global bending formula~\eqref{eq:eGaussianCurvature} and quantify properties intrinsic to spacetime.

\begin{table}
\caption{Symbols as in Figure~\ref{fig:gdp1} and their interpretations. All the vertices and angles are clearly labeled in the figure with sufficient detail.}
\vspace{0.2cm}
\centerline{
\begin{tabular}{c|cc}
\hline
~Symbol~&~Meaning/definition~&~Remarks~\\
\hline
\hline
$\gamma_k$&vertex at arc length $\lambda_k$&$k=0, \dots , \aleph$\\
\hline
$\gamma_{k^{\ast}}$&end point of some line segment chosen&$k^{\ast}\in\{i,j,l,m\}$\\
\hline
$\gamma_{k^{\prime}}$&other vertices of $\partial D_{\mathrm{u}}$&$k^{\prime}\notin\{i,j,l,m\}$\\
\hline
$\gamma_{k^{\prime\prime}}$&other vertices of $\partial D_{\mathrm{d}}$&$k^{\prime\prime}\notin\{ i,j,l,m,k^{\prime}\}$\\
\hline
\hline
$C_{pq}$&\makecell{line segment \\connecting any two vertices $\gamma_{p}$ and $\gamma_{q}$ on $\partial{\mathring D}$}&\makecell{$p,q=0, \dots , \aleph$\\ ($p\neq q$)}\\
\hline
\hline
${\alpha}_{k}$ (${\beta}_{k}$)&external (interior) angle of $\partial{\mathring D}$ at $\gamma_{k}$&${\alpha}_{k}+\beta_k=\pi$\\
\hline
${\alpha}^{\rm o}_{k}$ ($\tilde{\beta}^{\rm o}_{k}$)&external (interior) angle of $\partial D_{\mathrm{u}}$ at $\gamma_{k}$&${\alpha}^{\rm o}_{k}+\tilde{\beta}^{\rm o}_{k}=\pi$, $k=i,j,l,m$ \\
\hline
$\tilde{\alpha}^{\rm o}_{k}$ (${\beta}^{\rm o}_{k}$)&external (interior) angle of $\partial D_{\mathrm{d}}$ at $\gamma_{k}$&$\tilde{\alpha}^{\rm o}_{k}+{\beta}^{\rm o}_{k}=\pi$, $k=i,j,l,m$ \\
\hline
\hline
$D_{\mathrm{c}}$&sub-region may containing singularities&\\
\hline
$D_{\mathrm{u}}$/$D_{\mathrm{d}}$& sub-regions after cutting geometric ``hole''&\\
\hline
$D$&region before removing singular region $D_{\mathrm{c}}$&~$D=D_{\mathrm u}\cup D_{\mathrm d}\cup D_{\mathrm c}$~\\
\hline
$\mathring{D}$&region after removing $D_{\mathrm{c}}$&$\mathring{D}=D_{\mathrm u}\cup D_{\mathrm d}$\\
\hline
\end{tabular}
}
\label{tab:symbol}
\end{table}

\subsection{Measurement formula}
\label{sec:2.2} 
However, we still lack knowledge on how to determine the global bending measure. Indeed, we have to establish a formula for the measurement of $\mathring{\alpha}_{M}$.
Prior to this, we need to learn how to apply the global Gauss-Bonnet theorem over a singular region $D$.
If the lensing patch $D$ is a region with $h$ singularities, then we can remove $h$ singular sub-regions $D^{r}_{\rm c}$, $r=0, ... , h-1$, and obtain a non-singular region $\mathring{D}$ with $h$ geometric ``holes'', similar to the approach taken for $D_{\rm c}$ mentioned earlier. Consequently, we can apply the global Gauss-Bonnet theorem over the singular region $D$ after removing its singularities and creating various ``holes''.

\noindent\\
{\bf GLOBAL GAUSS-BONNET THEOREM.} {\it Let $\mathring{D}$ be a regular region of an oriented surface with $h$ geometric ``holes'', associated with $D^{r}_{\rm c}$, $r=0, ... , h-1$, as well as $h+1$ corresponding topological boundaries, 
such as $\partial{D}^{r}_{\rm c}, r=0, 1,~...~,h-1$, and $\partial{D}$. Then,
\begin{eqnarray}
\label{eq:gGBF0}
\begin{array}{rcl}
\displaystyle
\nonumber
\sum_{k}\alpha_{k}+\sum_{k}\int^{\lambda_{k+1}}_{\lambda_{k}}\kappa_{g}\,\mathrm{d} \lambda+\int\!\!\!\int_{\mathring{D}}\,K\,\mathrm{d} \sigma=2\pi\,\chi(\mathring{D}),~~~~~~
\hspace*{0mm}
\end{array}
\end{eqnarray}
where $\chi=\chi(\mathring{D})$ is the Euler characteristic number, $\lambda$ is the arc length, and $\kappa_{g}=\kappa_{g}\left(\lambda\right)$ is the geodesic curvature of the regular arcs of $\gamma=\partial{\mathring{D}}$. As shown in Figure~\ref{fig:gdp1}, $\gamma$ is positively oriented, parametrised by arc length $\lambda$. Generally, $\chi=1-2 g-h$, where $g$ is the genus of $D$~\cite{Nakahara2003}.  Please see Appendix \ref{app:T} for further details.}\\

As indicated by the bending formula~\eqref{eq:eGaussianCurvature}, this generalised measure is equal to the negative of the surface integral of the Gaussian curvature over $\mathring{D}$. Then, by the global Gauss-Bonnet theorem, the Gaussian bending measure $\mathring{\alpha}_{M}$ can be reexpressed in the following form, 
\begin{eqnarray}
\label{eq:GDA2}
\mathring{\alpha}_{M}=\sum_{k}\alpha_{k}-2\pi\,\chi(\mathring{D})+\Delta_{g},~
\hspace*{0mm}
\end{eqnarray}
where $\Delta_{g}$ is the correction to the Gaussian bending measure from the geodesic curvature $\kappa_{g}$ and occurs in the case that the light rays or other messengers do not move along geodesics due to other types of forces than gravity, given as
\begin{eqnarray}
\label{eq:GDA2add}
\nonumber
\Delta_{g}&=&\sum_{k}\int^{\lambda_{k+1}}_{\lambda_{k}}\kappa_{g}\,\mathrm{d} \lambda\\
\nonumber
&=&\oint_{\partial{D}}\kappa_{g}\,\mathrm{d} \lambda-\sum_{r}\oint_{\partial{D}^{r}_{\rm c}}\kappa_{g}\,\mathrm{d} \lambda\\
&=&\oint_{\partial{\mathring D}}\kappa_{g}\,\mathrm{d} \lambda, 
\end{eqnarray}
where $\partial{D}$ and $\partial{D}^{r}_{\rm c}$, $r=0, ... , h-1$, are all simple, smooth, closed curves, as required by the Gauss-Bonnet theorem in differential geometry~\cite{Carmo16,Chern00}. Here, $\partial{\mathring D}$ denote the boundary set of the region $\mathring{D}$, i.e., $\partial{\mathring D}=\partial{D}-\sum_{r}\partial{D}^{r}_{\rm c}$, where $\partial{D}$ will be referred to as the outer boundary of the singularity-free region $\mathring{D}\subset\Sigma$, and $\partial{D}^{r}_{\rm c}$ as the $r$-th inner boundary hereafter. The formula~\eqref{eq:GDA2} is indeed used for the measurement of the global bending measure.

\subsection{Further results and comments}
\label{sec:2.3} 

In physics, all the inner and outer boundaries should consist of a finite number of physically simple geodesic segments. 
As shown in equation~\eqref{eq:eGaussianCurvature}, the Gaussian bending measure is equivalent to the nagative of the surface integral of the Gaussian curvature over the singularity-free lensing patch $\mathring{D}$ with geometric ``holes''. 
Interestingly, the $r$-th inner boundary $\partial{D}^{r}_{\rm c}$ of $\mathring{D}$ can be chosen to approach infinitely close to the intrinsic boundary of $\Sigma$ surrounding its $r$-th singularity; an explicit example of the intrinsic boundary is the outer event horizon of a Kerr black hole. 
Accordingly, each inner boundary can be ideally defined as a corresponding intrinsic boundary. 
If so, the Gaussian bending measure $\mathring{\alpha}_{M}$ is termed as an absolute measure; 
otherwise, it is called a relative measure. 
In math, the absolute measure can be regarded as a relative measure. 
In either case, $\mathring{\alpha}_{M}$ is equivalent to the surface integral of the Gaussian curvature over $\mathring{D}$ rather than $D$. 
Hence, the relationship~\eqref{eq:GDA2} is established based on the area $\mathring{D}$ between the outer boundary and every intrinsic inner boundary
and is totally unrelated to the region enclosed within each intrinsic boundary. 

For instance, in the equatorial plane of a Kerr black hole, the sum of the external angles of the outer boundary $\partial{D}$ is completely independent of the region inside the outer event horizon of the Kerr black hole. In fact, one always has $\kappa_{g}\equiv0$ at the event horizon. Additionally, the external angles of the intrinsic inner boundary of $\mathring{D}$ at the event horizon disappear, denoted by $\alpha^{\rm I}_{k}=0$ for each of them. 
Thus, there are no extra corrections to the sum of the external angles of the outer boundary $\partial{D}$ from the intrinsic inner boundary, 
with all the contributions to this sum originating from the region between the inner and outer boundaries. In other words, the gravity within a Kerr black hole does not contribute the absolute bending measure.

In the following, we will proceed with our analysis using the absolute measure, although a similar analysis can be carried out with the relative measure; 
specifically when the Gaussian bending measure is mentioned without emphasis, it refers to the absolute measure. 
In current theories of gravity, there always exist intrinsic inner boundaries with $\Delta_{g}=0$ and $\sum_{k}\alpha^{\rm I}_{k}=0$, similar to those of a Kerr black hole at its outer event horizon. From now on, these intrinsic boundaries\footnote{If $\Delta_{g}\neq0$ and $\sum_{k}\alpha^{\rm I}_{k}\neq0$, we just need to make fixed corrections to them, as they are exclusively determined by the unique metric of the spacetime being investigated.} will be chosen as the inner boundaries, 
while the outer boundary should consist of a finite number of physically simple geodesic segments without any self-intersections, as we will demonstrate in the following sections. 
Obviously a necessary condition for the outer boundary $\partial{D}$ to be a geodesic curve is $\kappa_{g}\equiv0$. 
In a realistic event of gravitational bending, both massless and massive messengers move along geodesics. 
Thus, one has $\Delta_{g}=0$ in general for these messengers. 
Nevertheless, if these messengers are subject to any non-gravitational force, the effects of $\Delta_{g}$ must be considered. 

In summary, starting from the local version of the Gauss-Bonnet theorem, we have established the global theory, including the definition~\eqref{eq:edefinition} of the Gaussian bending measure, its global bending formula~\eqref{eq:eGaussianCurvature}, its relationship with external angles~\eqref{eq:GDA2}, its irrelevance to the region enclosed within each intrinsic boundary, and its methods of measurement, which is applicable to any type of messengers in the spacetime region $D\subset\Sigma$ with singularities.

\section{Applications of the global theory in stationary spacetimes}
\label{sec:3} 

In this section, we will investigate the global theory of Gaussian bending in a general stationary spacetime, 
and explore its potential applications in understanding diverse astrophysical phenomena and designing various experiments. 

\subsection{General analysis}
\label{sec:3.1} 

In GR, the metric of any stationary spacetime can be written in the general form, 
\begin{eqnarray}
\label{eq:StatioM}
\begin{array}{rcl}
\displaystyle
\mathrm{d} s^2 &=&g_{\mu\nu}\,\mathrm{d} x^{\mu}\,\mathrm{d} x^{\nu}\\[3mm]
&=&g_{00}\,\mathrm{d} t^2+g_{0i}\,\mathrm{d} t \,\mathrm{d} x^{i}+g_{i0}\,\mathrm{d} x^{i}\, \mathrm{d} t+g_{ij}\,\mathrm{d} x^{i} \mathrm{d} x^{j}, 
\end{array}
\end{eqnarray}
with $\mu,\nu\in\{0, 1,2,3\}$ and $i,j\in\{1,2,3\}$, where $x^{0}$ ($=\!t$) and $x^{i}$ (or $x^{j}$) represent the time-like and space-like coordinates, respectively. Thus, $g_{00}<0$. In GR, we always have $g_{\mu\nu}=g_{\nu\mu}$. 
For any given $t$, we can obtain a three-dimensional hypersurface, associated with a spatial line element, 
\begin{eqnarray}
\label{eq:SpaceLineE}
\begin{array}{rcl}
\displaystyle
\mathrm{d} \sigma^2 \,=\,g_{ij}\,\mathrm{d} x^{i}\mathrm{d} x^{j}. 
\end{array}
\end{eqnarray}
Then, within the hypersurface, a two-dimensional physical surface $\Sigma$ can always be chosen at our convenience, as demonstrated in~\cite{Zhang:2021ygh}. 
Let us assume that the physical surface $\Sigma$ is parametrised by
$$ \Sigma:~~ \left\{
\begin{aligned}
~x^{1} & = & x^{1}\left(u, \upsilon\right) \\
~x^{2} & = & x^{2}\left(u, \upsilon\right) \\
~x^{3} & = & x^{3}\left(u, \upsilon\right)
\end{aligned}
\right.~~~,
$$
where $\left(u, \upsilon\right)$ represent an alternative set of spatial coordinates. 
We then use $u^{i}~\left(i=1, 2\right)$ to denote $\left(u, \upsilon\right)$. Here, we can have $\{u^{i}\}\subset\{x^{i}\}$ or their functions. 
Thus, 
\begin{eqnarray}
\label{eq:MetricTransf}
\begin{array}{rcl}
\displaystyle
\mathrm{d} x^{i}=\frac{\partial x^{i}}{\partial u^{l^{\prime}}}\,\mathrm{d} u^{l^{\prime}},~~\mathrm{d} x^{j}=\frac{\partial x^{j}}{\partial u^{m^{\prime}}}\,\mathrm{d} u^{m^{\prime}}, 
\end{array}
\end{eqnarray}
where $l^{\prime}, m^{\prime}=1,2$. 
Substituting~\eqref{eq:MetricTransf} into~\eqref{eq:SpaceLineE} gives
\begin{eqnarray}
\label{eq:Sigma}
\mathrm{d} \sigma^{2}&=&E\,\mathrm{d}\,\!u^{2}+2 F\,\mathrm{d}\,\!u\,\mathrm{d}\upsilon+G\,\mathrm{d}\upsilon^{2}, 
\end{eqnarray}
where $E$, $F$, and $G$ are given by
\begin{eqnarray}
\label{eq:EFG0}
\nonumber
E&=&\frac{\partial x^{i}}{\partial u}\frac{\partial x^{j}}{\partial u}\,g_{ij}, \\
\nonumber
F&=&\frac{\partial x^{i}}{\partial u}\frac{\partial x^{j}}{\partial \upsilon}\,g_{ij}, \\
\nonumber
G&=&\frac{\partial x^{i}}{\partial \upsilon}\frac{\partial x^{j}}{\partial \upsilon}\,g_{ij}, 
\end{eqnarray}
where $F$ can be set as zero. 
Indeed, through coordinate transformations, the line element $\mathrm{d} \sigma^2$ can be rewritten in an orthogonal form, i.e., $F=0$~\cite{Carmo16}. 
Generally, the physical surface $\Sigma$ may have intrinsic inner boundaries around the singularities. 
In orthogonal coordinates, one can always set $E>0$ and $G>0$, 
as the singular region within each intrinsic inner boundary has no contribution to the Gaussian bending measure $\mathring{\alpha}_{M}$ (Section~\ref{sec:2}). 
Thus, when studying effects on $\mathring{\alpha}_{M}$, the physical surface $\Sigma$ should be limited by these intrinsic boundaries, 
which are determined by the boundary conditions of $E$ and $G$, especially as $E$ and $G$ approach their limits of zero or infinity. 
Within these boundaries, the Gaussian curvature $K$ can be simply expressed as
\begin{eqnarray}
\label{eq:GaussK}
\begin{array}{rcl}
\displaystyle
K=-\frac{1}{\sqrt{EG}}\left(\left(\frac{(\sqrt{E})_{\upsilon}}{\sqrt{G}}\right)_{\!\upsilon} +\left(\frac{(\sqrt{G})_{u}}{\sqrt{E}}\right)_{\!u}\right), 
\end{array}
\end{eqnarray}
where $X_{u}$ ($X_{\upsilon}$) denotes the partial derivative of the function $X=E,~G$ with respect to $u$ ($\upsilon$). 
In the lensing patch $D$ with singularities, the total curvature $K_{\rm tot}$ can be defined as
\begin{eqnarray}
\label{eq:total0}
\nonumber
K_{\rm tot}=\int\!\!\!\int_{\mathring{D}}\,K\mathrm{d} \sigma, 
\hspace*{0mm}
\end{eqnarray}
where $\mathring{D}=D$ in the absence of singularities. 
The definition of $K_{\rm tot}$ is a natural and sensible generalisation of the total curvature in differential geometry~\cite{Zhang:2021ygh}. 
This total curvature measures the basic properties of the global geometry of the lensing patch $D$. 
Then, it follows that $\mathring{\alpha}_{M}$ can be described by the following form, 
\begin{eqnarray}
\label{eq:total}
\nonumber
\mathring{\alpha}_{M}&=&-K_{\rm tot}\\ 
\nonumber
&=&\int\!\!\!\int_{\mathring{D}}\,\left(\left(\frac{(\sqrt{E})_{\upsilon}}{\sqrt{G}}\right)_{\!\upsilon} +\left(\frac{(\sqrt{G})_{u}}{\sqrt{E}}\right)_{\!u}\right)\,\rm{d}{u}\,\rm{d}{\upsilon}\\
&=&\oint_{\partial{\mathring{D}}}\left(-\frac{(\sqrt{E})_{\upsilon}}{\sqrt{G}}\,\rm{d}{u}+\frac{(\sqrt{G})_{u}}{\sqrt{E}}\,\rm{d}{\upsilon}\right), 
\hspace*{0mm}
\end{eqnarray}
where $\partial{\mathring{D}}$ is the boundary set of $\mathring{D}$ (Section~\ref{sec:2}). 
The first line is simply the definition~\eqref{eq:eGaussianCurvature}. 
The second line comes directly from equation~\eqref{eq:GaussK}. 
The last line uses Green's theorem. 
Here, the boundary set $\partial{\mathring{D}}$ is comprised of an outer boundary and at least one inner boundary, with each singularity being encircled by a unique inner boundary. 
As we will demonstrate below, such as in the equatorial plane of a Kerr black hole, 
these intrinsic inner boundaries are determined by the outer event horizon of the black hole within the lensing patch $D$. 
By contrast, in the singularity-free situation, there will be no intrinsic inner boundaries.

\subsection{Specific examples of applications}
\label{sec:3.2}

If the lensing patch $D\subset\Sigma$ contains singularities, 
we could cut ``holes'' along the intrinsic inner boundaries of the physical surface $\Sigma$, encircling these singularities, similar to what was done in the previous section, 
thus forming a singularity-free region $\mathring{D}$. 
On the other hand, the outer boundary of $\mathring{D}$ should be composed of physically simple geodesic segments. 
In differential geometry, it is actually the boundary of a geodesic {\it polygon}, such as a geodesic digon or a geodesic triangle. 
Interestingly, if there is only one geometric ``hole'' in the singularity-free region $\mathring{D}\subset\Sigma$, we find
\begin{eqnarray}
\label{GDAob}
\nonumber
\mathring{\alpha}_{M}&=&-K_{\rm tot}\\
&=&\oint_{\partial{D}}\left(-\frac{(\sqrt{E})_{\upsilon}}{\sqrt{G}}\,\rm{d}{u}+\frac{(\sqrt{G})_{u}}{\sqrt{E}}\,\rm{d}{\upsilon}\right)+\mathfrak{Z}_{0},~~~
\end{eqnarray}
where the lensing patch $D\subset\Sigma$ is already orthogonally parametrised in the coordinates $(u,\upsilon)$, and the boundary $\partial{D}$ can be chosen at our convenience. 
Note that this formula holds true if and only if the Gauss-Bonnet theorem is applicable; the derivations and interpretations are provided in Appendix~\ref{app:A}. 
Here, the constant $\mathfrak{Z}_{0}$ is, in fact, a topological invariant. Exactly, its value can be determined through calculations over an outer boundary $\partial{D}$ selected for our case. An instance of this is the equatorial plane of a Kerr black hole, where we can choose an outer boundary with its vertices located at spatial infinity; see Appendix~\ref{app:A} for further details. Then, because of the asymptotical flatness of the Kerr spacetime, one has $\mathfrak{Z}_{0}=0$ on this outer boundary. In this case, there is only one intrinsic inner boundary. As indicated clearly in the formula~\eqref{GDAob}, the Gaussian deflection measure $\mathring{\alpha}_{M}$ cannot be influenced by the inner boundary, or its value is only dependent of the outer boundary. This is an intriguing result, and it holds practical significance. For example, the use of this formula will greatly simplify our comprehension of the physics related to the Gaussian bending measure.

\begin{figure}
\centering
\includegraphics[width=0.85\columnwidth]{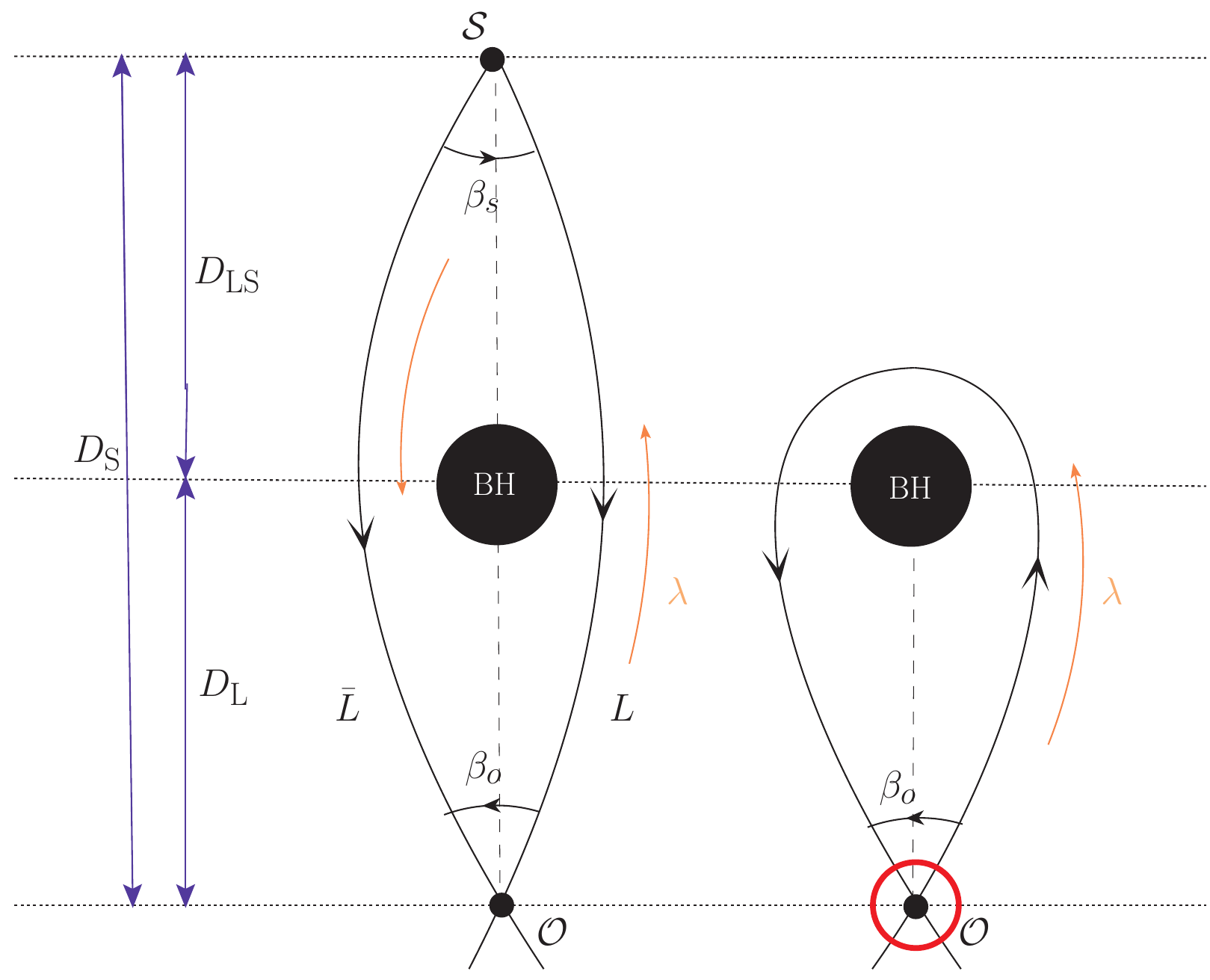}\\[1mm]
\caption{
Geodesic digon and monogon: the geometry of global Gaussian bending from a side-on view. The left panel shows that the light rays emitted from the source at the point  $\mathcal{S}$ propagate along the geodesic curves $L$ and $\bar{L}$, respectively, pass by a black hole, and ultimately reach the observer at the point $\mathcal{O}$, forming a geodesic digon. The right panel illustrates a geodesic monogon, along the outer boundary of which the light rays originating from the point $\mathcal{O}$ propagate, are then bent by a black hole, finally arrive back at the point $\mathcal{O}$, i.e., $\mathcal{S}=\mathcal{O}$. As marked by the red circle, there exists a self-intersection at the vertex of the geodesic monogon. 
}
\label{fig:digon}
\end{figure}

In general theories of gravity, a closed inner boundary can always be found in the surrounding of a singularity on the physical surface $\Sigma$. 
Especially for a black hole, its event horizon can be seen as an intrinsic inner boundary $\partial{D}_{\rm c}$, along which the geometric ``hole'' is pre-defined by the specific theory. 
In such a case, the inner boundary is determined solely by the black hole spacetime itself, naturally forming a geometric ``hole'' on the physical surface $\Sigma$. 
In GR, such an intrinsic inner boundary can always be found with $\kappa_{g}\equiv0$, such as that of a Kerr black hole at the outer event horizon, and it has no external angles, i.e., $\alpha^{\rm I}_{k}=0$ for each $k$. 
As depicted in the left panel in Figure~\ref{fig:digon}, in the case of an Einstein ring, 
the light rays passing by a black hole from a source to us follow two geodesic trajectories denoted as $L$ and $\bar{L}$. 
Thus, one has $\Delta_{g}=0$ along these trajectories. 
In this case, the lensing patch $D$ is enclosed by the two light trajectories, with a singularity marked by the black hole. 
The two interior angles of the outer boundary, particularly denoted as $\beta^{\rm E}_{s}$ and $\beta^{\rm E}_{o}$, are actually the measured intersection angles by the local observers between the two light trajectories at the points $\mathcal{S}$ and $\mathcal{O}$, respectively. 
Evidently, it is the global bending measure $\mathring{\alpha}_{M}$ in equation~\eqref{eq:eGaussianCurvature} that connects the two interior angles, rather than $\alpha_{M}$ in equation~\eqref{eq:GaussianCurvature}. 
From equation~\eqref{eq:GDA2}, one obtains
\begin{eqnarray}
\label{eq:OA1}
2\pi-\sum_{k=s,o}\alpha^{\rm E}_{k}=\beta^{\rm E}_{s}+\beta^{\rm E}_{o}=2\pi\left(2 g+h\right)+K_{\rm tot},~~~
\hspace*{0mm}
\end{eqnarray}
where $\alpha^{\rm E}_{k}$ represents the $k$-th external angle of the outer boundary. 
This establishes a link between the two interior angles and the global properties of the geometry and topology of the lensing patch. 
Here, $\left(g,h\right)=\left(0,1\right)$. 
Besides, such scenarios can also been observed in the shadow imaging of a black hole, 
where the photons from background sources pass near the black hole, while the boundary of the shadow of the black hole precisely traces the radius of the outer boundary of the photon region~\cite{Qiao2022}.
In this case, the angle $\beta^{\rm E}_{s}$ at the point $\mathcal{S}$ can be assumed to be zero if the light rays originate at spatial infinity. 
Thus, one has
\begin{eqnarray}
\label{eq:OA3}
\beta^{\rm E}_{o}=2\pi\left(2 g+h\right)+K_{\rm tot},~
\hspace*{0mm}
\end{eqnarray}
which is derived directly from equation~\eqref{eq:OA1}.
If the lens is a normal object like an isolated neutron star, $\left(g,h\right)=\left(0,0\right)$. 
Then, from~\eqref{eq:OA1}, one obtains
\begin{eqnarray}
\label{eq:OA0}
\beta^{\rm E}_{s}+\beta^{\rm E}_{o}=\int\!\!\!\int_{D}\,K\,\mathrm{d} \sigma,~
\hspace*{0mm}
\end{eqnarray}
where $D$ is singularity-free. The right-hand side is indeed the total curvature $K_{\rm tot}$. 
In this case, $\mathring{D}=D$. 
Therefore, the global bending formula presented here enables us to establish a connection between the properties of the lens, the source, and the particles as messengers to the observables. In astronomy, this may have direct applications. For instance, in a gravitationally lensed $\gamma$-ray burst (GRB)~\cite{Paynter:2021wmb,Wang:2021ens,Lin:2021hae}, the opening angle $\beta^{\rm E}_{s}$ of the GRB jet can be determined from the GRB afterglow, $\beta^{\rm E}_{o}$ can be observationally limited by high-energy telescopes, and the arc lengths related to $L$ and $\bar{L}$ of $\partial{D}$ can be constrained by the time delay between the two GRB images. In general, the mass $M$ of the lens is in the range of $\sim10^{4}-10^{6}~M_{\odot}$, and thus, the lens is an intermediate-mass black hole, where $M_{\odot}$ is the mass of the Sun~\cite{Paynter:2021wmb,Wang:2021ens,Lin:2021hae}. By combining equation~\eqref{eq:OA1} with current observations, we can put additional constraints on the physics of GRB jets. In the near future, similar applications can also be expected in an Einstein ring and in the shadow imaging of a black hole.

\begin{figure}
\centering
\includegraphics[width=0.85\columnwidth]{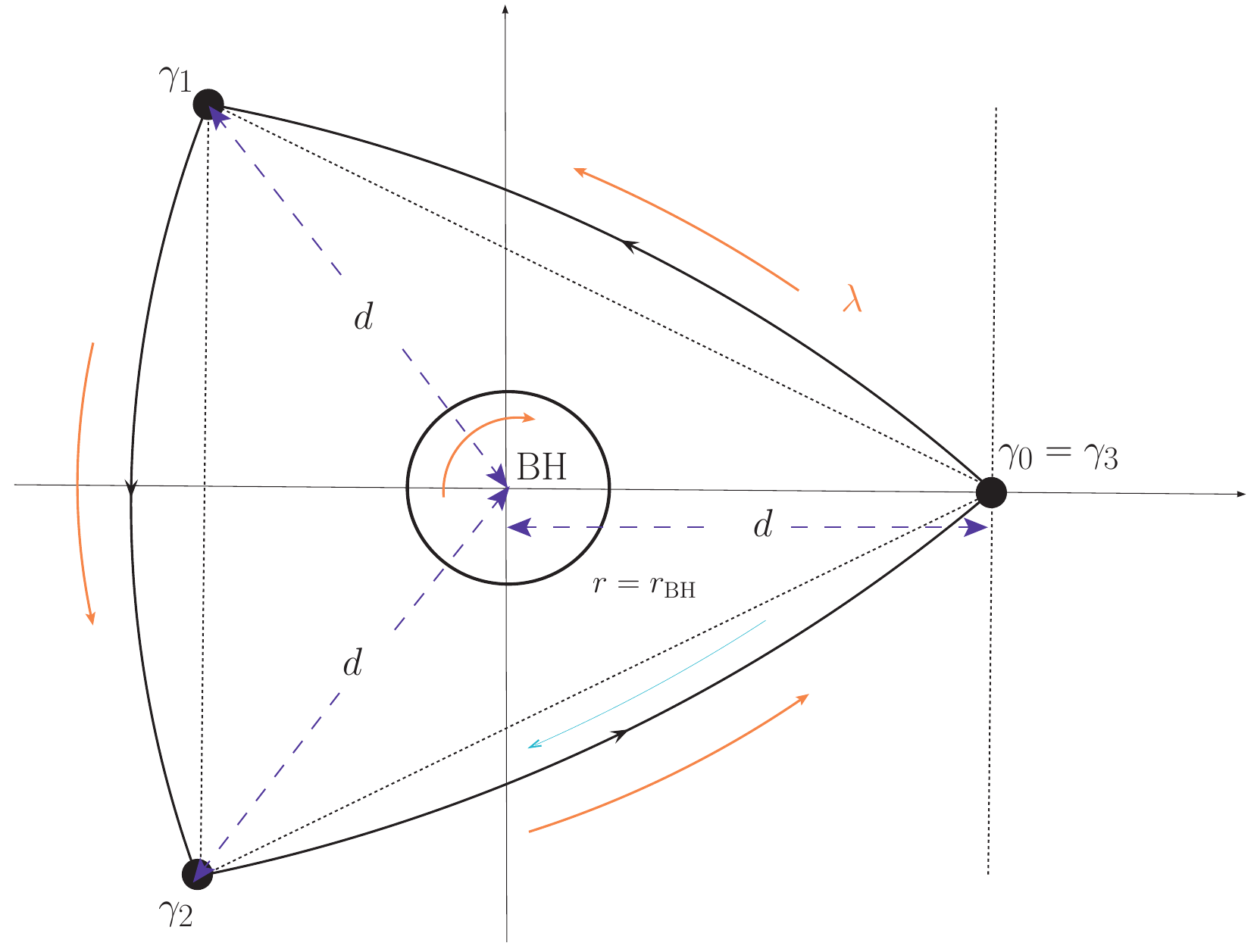}
\caption{
Geodesic triangle. At the three vertices $\gamma_{k},~k=1,2,3$, there are three devices (or mirrors) that function as both detectors and emitters of photons. Here, $d$ denotes the distance of each vertex to the center of mass of the black hole. The beamed light rays from these devices may travel along the outer boundary of the geodesic triangle, establishing a physically simple, closed loop of photons. 
}
\label{fig:triangle}
\end{figure}

Furthermore, based on the global theory, we can design experiments and develop strategies to test predictions within and beyond GR. 
For instance, when dealing with a geodesic {\textit triangle}, we have 
\begin{eqnarray}
\label{eq:triangle}
2\pi-\sum_{k=1}^{3}\alpha^{\rm E}_{k}=\sum_{k=1}^{3}\beta^{\rm E}_{k}-\pi=2\pi\left(2 g+h\right)+K_{\rm tot},~
\hspace*{0mm}
\end{eqnarray}
where $\alpha^{\rm E}_{k}$ ($\beta^{\rm E}_{k}$) denotes the $k$-th external (interior) angle of the outer boundary. 
It asserts that the excess of 2$\pi$ over the sum of external angles of the geodesic triangle (or the excess over $\pi$ of the sum of interior angles) is completely determined by the topological and geometrical properties of the lensing patch $\mathring{D}$. As shown by this equation, the change in $\left(g,h,K_{\rm tot}\right)$ could be used as an indicator for detecting the presence and number of singularities in certain lensing patches.
Accordingly, we can conduct some experiments to detect the presence of invisible objects like primordial black holes (PBHs)\footnote{Unlike the black holes that are formed from the collapse of massive stars, PBHs are thought to have originated from the extreme density fluctuations in the early universe. They could have a wide range of masses, from microscopic to supermassive. Generally, these black holes may have a mass large enough that Hawking radiation is insufficient to cause them to evaporate, with a mass estimated to be $\gtrsim10^8$ kg (or $\gtrsim10^{-22}~M_{\odot}$), and they have not yet been observed. } with the mass of an asteroid or the Earth in our Solar system. 
For example, we could position three devices (or mirrors) to serve as both detectors and emitters of photons at the three vertices of the geodesic triangle, as illustrated in Figure~\ref{fig:triangle}. We would then enable them to receive and emit light rays, forming the Gaussian bending of light on a geodesic triangle patch. 
If a black hole were to pass through the geodesic triangle, we might detect this event through the change in $\left(g,h,K_{\rm tot}\right)$, where $K_{\rm tot}$ is closely related to the spin and mass of the black hole. 
This would enable us to directly extract basic information about the black hole by measuring the sum of external angles of the polygon or the Gaussian bending measure, as detailed in equation~\eqref{eq:GDA2}, which provides further details important for practical experiments. Therefore,  the global Gaussian bending measure of massless or massive messengers can serve to probe the intrinsic properties of spacetime though the intersection angles between their trajectories.

However, if the light rays are emitted at a point $\mathcal{O}$, then bent by a black hole, and ultimately return to the starting point $\mathcal{O}$, i.e., $\mathcal{S}=\mathcal{O}$, a closed light trajectory can be formed, as illustrated by the geodesic ${\textit monogon}$ in the right panel of Figure~\ref{fig:digon}. 
A self-intersection subsequently occurs at one point on the closed null geodesic, specifically at the vertex of the geodesic monogon, 
as marked by the red circle. 
Consequently, the closed null geodesic is no longer mathematically simple. In other words, the first-order tangential derivative of the geodesic takes two different values at one single point, even though this first-order derivative is completely determined by one single differential equation, directly violating the condition of the uniqueness of the solution to the differential equation~\cite{Carmo16,Chern00}. 
This clearly indicates that the first-order derivative is no longer continuous at the vertex. 
It is important to note that the Gauss-Bonnet theorem pertains to the second-order derivative. 
For example, the geodesic curvature $\kappa_{g}$ is defined based on the second-order tangential derivative~\cite{Carmo16,Chern00}. 
Thus, the vertex becomes singular at second order. 
As a result, the presence of self-intersection violates the conditions required by the Gauss-Bonnet theorem, namely the continuity of the first-order derivative and the uniqueness of the solution to an differential equation, thus resulting in the failure of the Gauss-Bonnet theorem in describing the relationship between the external (or interior) angles and the total geodesic and Gaussian curvatures, a topic that is far beyond the scope of this work.

\section{Spacetime effects from rotating black holes}  
\label{sec:4} 

In this section, we will investigate how a rotating black hole affects the global Gaussian bending measure of massless or massive messengers, including photons, neutrinos, cosmic rays, and gravitational waves. Generally, this study can be carried out on any physical surface $\Sigma$, defined by the line element~\eqref{eq:Sigma}.

\subsection{The Kerr case as an example}
\label{sec:4.1}

As an example, let us work it out in the equatorial plane of a Kerr black hole as the physical surface $\Sigma$ chosen. 
In Boyer-Lindquist coordinates $\left(t,r,\phi\right)$, the metric of this plane is given by~\cite{Weinberg:1972kfs,Chandrasekhar1983,Rindler2006book,Carroll2014}
\begin{eqnarray}
\label{eq:Kerr}
\nonumber
\mathrm{d} s^2\!=\!&\!-\!&\!\left(1\!-\!\frac{2 M}{r}\right)\mathrm{d} t^2\!-\!\frac{4 a M}{r}\!\mathrm{d} t \,\mathrm{d}\phi\!+\!\left(1-\!\frac{2 M}{r}\!+\!\frac{a^2}{r^2}\right)^{-1}\!\mathrm{d} r^2\\
\nonumber
&\!+\!&\!\left(1+\!\frac{a^2}{r^2}\!+ \!\frac{2 M}{r}\frac{a^2}{r^2}\right)r^2\mathrm{d}\phi^2, \!\!\!\!\!\!\!\!\!\!
\end{eqnarray}
where $M$ is the mass of the black hole, and $a$ is the angular momentum per unit mass. 
As is custom in GR, the units are chosen such that $c=G=1$, where $c$ is the speed of light and $G$ is Newton's constant of gravitation. 
Then, the spatial line element of the equatorial plane can be reexpressed in the coordinates $\left(u, \upsilon\right)\!=\!\left(r, \phi\right)$ as
\begin{eqnarray}
\label{eq:Sigma_kerr}
\nonumber
\mathrm{d} \sigma^{2}&\!=\!&E\,\mathrm{d}\,\!u^{2}+2 F\,\mathrm{d}\,\!u\,\mathrm{d}\upsilon+G\,\mathrm{d}\upsilon^{2} \!\!\!\!\!\!\!\!\!\!\\[2mm]
\nonumber
&\!=\!&\left(1\!-\!\frac{2 M}{r}\!+\!\frac{a^2}{r^2}\right)^{-1}\!\mathrm{d} r^2\!+\!\left(1+\!\frac{a^2}{r^2}\!+ \!\frac{2 M}{r}\frac{a^2}{r^2}\right)r^2\mathrm{d}\phi^2, 
\end{eqnarray}
which is already orthogonally parametrised, i.e., $F=0$. Correspondingly, 
\begin{eqnarray}
\label{eq:EFG}
\begin{array}{rcl}
\displaystyle
~E\!=\!\left(1-\frac{2 M}{r}+\frac{a^2}{r^2}\right)^{-1},~G\!=\!\left(1+\!\frac{a^2}{r^2}\!+ \!\frac{2 M}{r}\frac{a^2}{r^2}\right)r^2, ~~~~ 
\end{array}
\end{eqnarray}
by which we can define a global Gaussian bending measure and establish its global theory. 
Next we focus on the spacetime regions that may affect the Gaussian bending measure. 
In fact, as seen from the observers outside an event horizon, if a surface $\Sigma$ has influence on gravitational bending measurements, 
it should be limited to the spatial regions defined by $E>0$ and $G>0$. 
To fully understand this in GR, let us consider test-particles. 
Once these particles enter the event horizon, i.e., $E<0$, they are classically unable to establish causal connections with external observers or particles of any type, 
which is directly resulted from the definition of event horizon, 
thereby having no impact on the measurements performed by these external observers. 
Therefore, the measured values by the external observers for the Gaussian bending measure cannot be influenced by what occurs within the event horizon. 
Actually, as demonstrated in equation~\eqref{GDAob}, these values depend only on the outer boundary chosen. 
Besides, in the case of $E<0$, the coordinate $r$ behaves more like time rather than space, the geometry determined by~\eqref{eq:Sigma_kerr} is no longer associated with a space-like region. Thus, the geometric quantity $K$ does not have the same meaning as the Gaussian curvature defined by C. F. Gauss over a space-like surface~\cite{Carmo16,Chern00}. From a measurement perspective, this geometry cannot contribute to the gravitational bending of particles traveling in a space-like region. 
Accordingly, in the Kerr equatorial plane, we need
\begin{eqnarray}
\label{eq:Horizon}
\begin{array}{rcl}
\displaystyle
r> r_{+}^{h}=\!M\!+\!\sqrt{M^2\!-\!a^2}, 
\end{array}
\end{eqnarray}
where $r_{+}^{h}$ is also the radius of the outer event horizon of the Kerr black hole at its equatorial plane. 
In terms of measurement, this imposes an extra boundary condition on the physical surface $\Sigma$. 
Mathematically, $r$ always tends towards $r_{+}^{h}$, as $E$ approaches infinity. 
According to the viewpoint of any external observer, the measurably meaningful intrinsic inner boundary of the lensing patch is determined by $r=r_{+}^{h}$, along which one obtains
\begin{eqnarray}
\label{eq:kgformula}
\begin{array}{rcl}
\displaystyle
\kappa_{g}=\frac{1}{2\sqrt{E}}\frac{\partial \,{\rm log} \,G}{\partial r}\bigg|_{r=r_{+}^{h}}=0, 
\end{array}
\end{eqnarray}
directly from the Liouville formula~\cite{Carmo16,Chern00}. Indeed, it holds well for any $\theta=const.$ in a Kerr case. 
Accordingly, we always have $\Delta_{g}=0$ along the intrinsic inner boundary at the outer event horizon. 
Notably, there are no external angles present at the intrinsic inner boundary, i.e., $\alpha^{\rm I}_{k}=0$. 
By combining equations~\eqref{eq:EFG} with~\eqref{eq:GaussK}, we can derive the Gaussian curvature as
\begin{eqnarray}
\label{eq:GaussK1}
\nonumber
K\!=\!-\!\frac{M}{r^3}\bigg[\frac{\!1\!+\!\frac{5 a^6 M}{r^7}\!+\!\frac{3 a^6\!-\!8 a^4 M^2}{r^6}\!-\!\frac{2 a^4 M}{r^5}\!+\!\frac{7 a^4}{r^4}\!-\!\frac{11 a^2 M}{r^3}\!+\!\frac{5 a^2}{r^2}}{1+\frac{4 a^4 M^2}{r^6}\!+\!\frac{4 a^4 M}{r^5}\!+\!\frac{a^4}{r^4}\!+\!\frac{4 a^2 M}{r^3}\!+\!\frac{2 a^2}{r^2}}\bigg],~
\end{eqnarray}
which recovers to the Schwarzschild result in the zero-spin limit $a\to0$. 
Likewise, by substituting equations~\eqref{eq:EFG} into equation~\eqref{eq:total}, the total Gaussian curvature $K_{\rm{tot}}$ can be rewritten as
\begin{eqnarray}
\label{eq:totalK}
K_{\rm tot}=-\oint_{\partial{\mathring{D}}}\frac{\sqrt{r^2-2 M r+a^2} \left(1-\frac{a^2 M}{r^3}\right)}{\sqrt{r^2+a^2\left(1+\frac{2 M}{r}\right)}}\mathrm{d}\phi, 
\hspace*{0mm}
\end{eqnarray}
where $r\!=\!r\left(\phi\right)$ is determined by the inner and outer boundaries of $\partial{\mathring{D}}$, denoted as $\partial{D}_{c}$ and $\partial{D}$, respectively. 
Here, $D_{c}$ represents the region with a singularity cut along the intrinsic inner boundary $\partial{D}_{c}$ of the equatorial plane $\Sigma$. 
Then, by the Gaussian bending formula~\eqref{eq:eGaussianCurvature}, the generalised bending measure $\mathring{\alpha}_{M}$ can be further expressed as
\begin{eqnarray}
\label{eq:total_kerr0}
\mathring{\alpha}_{M}&=&\oint_{\partial{\mathring{D}}}\frac{\sqrt{r^2-2 M r+a^2} \left(1-\frac{a^2 M}{r^3}\right)}{\sqrt{r^2+a^2\left(1+\frac{2 M}{r}\right)}}\mathrm{d}\phi\,\\[2.5mm]
&=&\left[~...~\right]_{\partial{D}}-\left[~...~\right]_{\partial{D}_{c}}\left(\equiv0\right)\\[2.5mm]
\label{eq:total_kerr}
&=&\oint_{\partial{D}}\frac{\sqrt{r^2-2 M r+a^2} \left(1-\frac{a^2 M}{r^3}\right)}{\sqrt{r^2+a^2\left(1+\frac{2 M}{r}\right)}}\frac{\mathrm{d} r~~}{\big[\frac{\mathrm{d} r}{\mathrm{d}\phi}\big]_{\gamma}}\,,~~~~~~~
\hspace*{0mm}
\end{eqnarray}
where $\left[~...~\right]_{\partial{D}_{c}}$ and $\left[~...~\right]_{\partial{D}}$ represent the line integral~\eqref{eq:total_kerr0} along the inner and outer boundaries, respectively. 
Here, $\big[\frac{\mathrm{d} r}{\mathrm{d}\phi}\big]_{\gamma}$ refers to $\frac{\mathrm{d} r}{\mathrm{d}\phi}$ as a function of $r$ along the closed, piecewise regular, parametrised outer boundary $\gamma=\partial{D}$, which is composed of a finite number of simple segments, and this function varies between different segments. 
Let us think about this line by line. The first line comes directly from substituting equation~\eqref{eq:totalK} into equation~\eqref{eq:eGaussianCurvature}. 
In the second line, the second term becomes zero along $\partial{D}_{c}$ at $r=r_{+}^{h}$. 
Once entering the horizon, $E$ may become negative, 
and the Gaussian curvature $K$ no longer contributes to the Gaussian deflection measure $\mathring{\alpha}_{M}$,
which aligns with our insight at the end of section~\ref{sec:2}.
The third line indicates that $\mathring{\alpha}_{M}$ is reliant on the outer boundary $\partial{D}$ rather than the inner boundary $\partial{D}_{c}$, 
further confirming the result from equation~\eqref{GDAob} in a more physically meaningful way. 
In a realistic event of gravitational bending, $\partial{\mathring{D}}$ should consist of simple geodesic segments. 
Thus, $\big[\frac{\mathrm{d} r}{\mathrm{d}\phi}\big]_{\gamma}$ can be derived from the geodesic differential equations. 
Note that the bending formula~\eqref{eq:total_kerr} is derived without any of further assumptions or approximations. 
Physically, it can be applied to describe the propagation of any type of messengers, such as gravitational waves, regardless of whether the gravitational interaction is massless or not.
From this formula~\eqref{eq:total_kerr}, we can further conclude that the rotation of the black hole does contribute to the Gaussian bending measure via the $a$-dependent term. 
Hence, it is highly possible to directly extract the information of a black hole, such as its $\left(a, M\right)$, by making precise measurements of the Gaussian bending measure of both the massless and massive messengers in the future.

\subsection{Analytical formulations along a chosen outer boundary} 
\label{sec:4.2} 

Currently, photons remain the most important messenger because of their unique properties, including being massless, traveling at the speed of light, and their ability to carry information about fundamental processes in the universe. In general, they follow null geodesics. 
Thus, the focus here is on the outer boundary $\partial{D}$ composed of null geodesic segments. 
Now we can compute the unknown term $\big[\frac{\mathrm{d} r}{\mathrm{d}\phi}\big]_{\gamma}$ involved in the bending formula~\eqref{eq:total_kerr}. 
This term is actually determined by the light orbital equation, 
which is given by~\cite{Hioki:2009na} as
\begin{eqnarray}
\label{eq:loe_KerrR}
\left(\frac{\mathrm{d} r}{\mathrm{d} \phi}\right)^2\!\!\!\!=\!\left(r^2\!-\!2 M r\!+\!a^2\right)^2\!\frac{ \big[\frac{2M}{r}\!\left(a\!-\!b\right)^2\!+\!\left(a^2\!-\!b^2\right)\! +\!r^2\big]}{\big[2\left(a\!-\!b\right)M\!+\!b\,r\big]^2},~~~
\end{eqnarray}
where $b$ is the impact parameter. 
The outer boundary $\partial{D}$ can be described in segments by corresponding light orbital equations, respectively. 
Note that any light orbit is essentially a geodesic~\cite{SS2018}. 
Thus, the outer boundary also meets the condition $\Delta_{g}=0$, just like the intrinsic inner boundary. 
Besides, there is only one geometric ``hole'' in the equatorial plane of a Kerr black hole, i.e., $\left(g,h\right)=\left(0,1\right)$. Thus, $\chi(\mathring{D})=0$. 
Then, if the the outer boundary is a geodesically polygonal, by formula~\eqref{eq:GDA2}, one gets
\begin{eqnarray}
\label{eq:GDA2-4}
\mathring{\alpha}_{M}=\sum_{k}\alpha_{k}=\sum_{k}\alpha^{\rm E}_{k}, 
\hspace*{0mm}
\end{eqnarray}
with $\sum_{k}\alpha^{\rm I}_{k}=0$, where $\alpha^{\rm E}_{k}$ denotes the $k$-th external angle of the outer boundary. 
In such a case, the generalised Gaussian bending measure $\mathring{\alpha}_{M}$ is equal to the sum of these external angles. 
It means that the global Gaussian bending measure can always be measured from the external angles of the outer boundary. 
If the physical surface $\Sigma$ is a flat plane, the sum of the external angles is 2$\pi$. 
So the excess of $2\pi$ over the sum of the external angles, or $2\pi\!-\!\mathring{\alpha}_{M}$, can be used to quantify the extent to which $\Sigma$ deviates from a flat plane.

Once the outer boundary $\partial{D}$ is given, we can establish the relationship of $2\pi\!-\!\mathring{\alpha}_{M}$ with the basic parameters of a black hole, such as its spin $a$ and mass $M$, as shown by the bending formula~\eqref{eq:total_kerr}. 
According to this relationship, we can design experiments to determine the Gaussian bending measure that involves geodesic polygons, 
with each side of these polygons satisfying a corresponding light orbital equation. 
The orbital equations are determined by the impact parameters and the distances of the vertices to the black hole. 
These impact parameters may vary between different sides, while the distances of the vertices to the black hole may also differ among these vertices. 
Then, by combining these light orbital equations with the bending formula~\eqref{eq:total_kerr}, we can calculate the spacetime effects of a rotating black hole on the global Gaussian bending measure of light. Conversely, we can further enhance these effects by choosing the outer boundary appropriately, which is of guidance in probing gravity beyond GR. For example, people investigated the local spacetime effects of dark energy on the propagation of light on a geodesic {\it tetragon} patch based on the local theory~\cite{Zhang:2021ygh}, and found that these effects can be easily enhanced by at least 14 orders of magnitude through the proper choice of the outer boundary.

\section{Prospects for applications in future experiments} 
\label{sec:5} 

\subsection{Comparison with other measures} 
\label{sec:5.1} 

The traditional theory of gravitational lensing~\cite{Refsdal1964,Bourassa1975,Narayan1992}, initially proposed by Einstein in 1936~\cite{Einstein:1936llh}, 
has been well developed and have successfully explained numerous astronomy observations. 
As displayed by the digon in Figure~\ref{fig:digon}, it generally involves the imaging by a gravitational lens~\cite{Lodge1919} of a light source, including the positions of the images, their separation, their time delay, their magnifications, and so on~\cite{Schneider2006,Rindler2006book}, which focuses on the group behaviors of a beam of light rays and is somewhat more than just the gravitational bending of light~\cite{Einstein:1911vc}. This theory is proposed based on the lens equation where the usual deflection angle plays a central role~\cite{Einstein:1936llh,VE2000,Refsdal1964,Bourassa1975,Narayan1992,Schneider2006}, rather than the Gaussian bending measure $\alpha_{M}$~\eqref{eq:GaussianCurvature} defined in~\cite{Zhang:2021ygh}  or $\mathring{\alpha}_{M}$~\eqref{eq:eGaussianCurvature} introduced in this work. 

Note, the usual deflection angle is defined as the Euclidean intersection angle between the outgoing light ray at the observer point far away from the lens and the incident light ray at the source point at spatial infinity~\cite{Carroll2014,Lake:2007dx,Zhang:2021ygh,RI2007}. In fact, the Euclidean intersection angle is just a coordinate angle, and it is dependent of the coordinate system chosen~\cite{Zhang:2021ygh,Lake:2007dx}. Thus, the usual deflection angle will deviate from the measurable intersection angle in curved spacetime regions~\cite{Lake:2007dx} and will become physically unmeasurable when the gravitational fields, even weak, cannot be ignored~\cite{RI2007,Ishak:2008ex,Arakida:2011ty}. 
Especially in a non-asymptotically flat curved spacetime, the usual deflection angle will become unmeasurable throughout the spacetime~\cite{Zhang:2021ygh}. To be physically meaningful, the usual deflection angle must be defined over the distant flat regions within an asymptotically flat spacetime. 
Take, for example, a gravitational lensing event. As illustrated in Figure~\ref{fig:digon}, we denote the distances from the source to the observer, from the source to the lens, and
from the lens to the observer as $D_{\rm S}$, $D_{\rm LS}$, and $D_{\rm L}$, respectively. 
If this lensing event can be accurately described by the traditional theory of gravitational lensing, 
these distances, $D_{\rm S}$, $D_{\rm LS}$, and $D_{\rm L}$, must be sufficiently large for the gravitational field to be weak enough, 
or else we cannot define the usual deflection angle effectively~\cite{Zhang:2021ygh}. 
Nevertheless, in the shadow imaging of a black hole, the light rays may originate from the immediate vicinity of the black hole~\cite{Hioki:2009na,Perlick:2021aok}. 
Thus, the traditional theory of gravitational lensing is no longer applicable in this case, because $D_{\rm LS}$ may be as small as $\sim3M$, and thus, the influence of the strong-gravity-induced curvature cannot be ignored. 
Now, the global theory of Gaussian bending has been established in a general stationary spacetime, allowing us to investigate the spacetime effects of the black hole on the bending of light in any strong-field regions. 

In addition to the Gaussian bending measure, there are other similar bending measures defined over non-singular regions in the shapes of triangles or tetragons, like those presented in the literature \cite{Arakida:2017hrm,Sanchez:2023ckq}. Unlike those measures, the Gaussian bending measure can be defined over different shapes of lensing regions, not limited to just non-singular triangles or tetragons. This flexibility allows us to design experiments in the lensing regions of different shapes.
In our approach, the lensing patch can take the shape of a digon, triangle, or other shapes, depending on our preference.
Besides, the outer boundary of the lensing patch can be chosen at our convenience.
For simplicity, we only consider lensing patches in the shapes of regular (or equilateral) geodesic polygons.

\subsection{Numerical analysis and further discussion} 
\label{sec:5.2} 

In this subsection, we still focus on the equatorial plane of a Kerr black hole. 
Denote by $d$ the distance of each vertex to the center of mass. 
Theoretically, it might be feasible to have a few devices orbiting a black hole in a circular orbit of radius $d$, and emitting light rays to nearby devices. As previously mentioned, these light rays would then be bent by the black hole and subsequently received by these devices, forming various geodesic polygons, such as digons and triangles. 
Here, to illustrate the strong-field effects more clearly and straightforwardly, we choose to calculate the generalised bending measure $\mathring{\alpha}_{M}$ over the lensing patch $D$ defined by a regular digon or a regular triangle. 

\begin{figure}
\centering
\includegraphics[width=1\columnwidth]{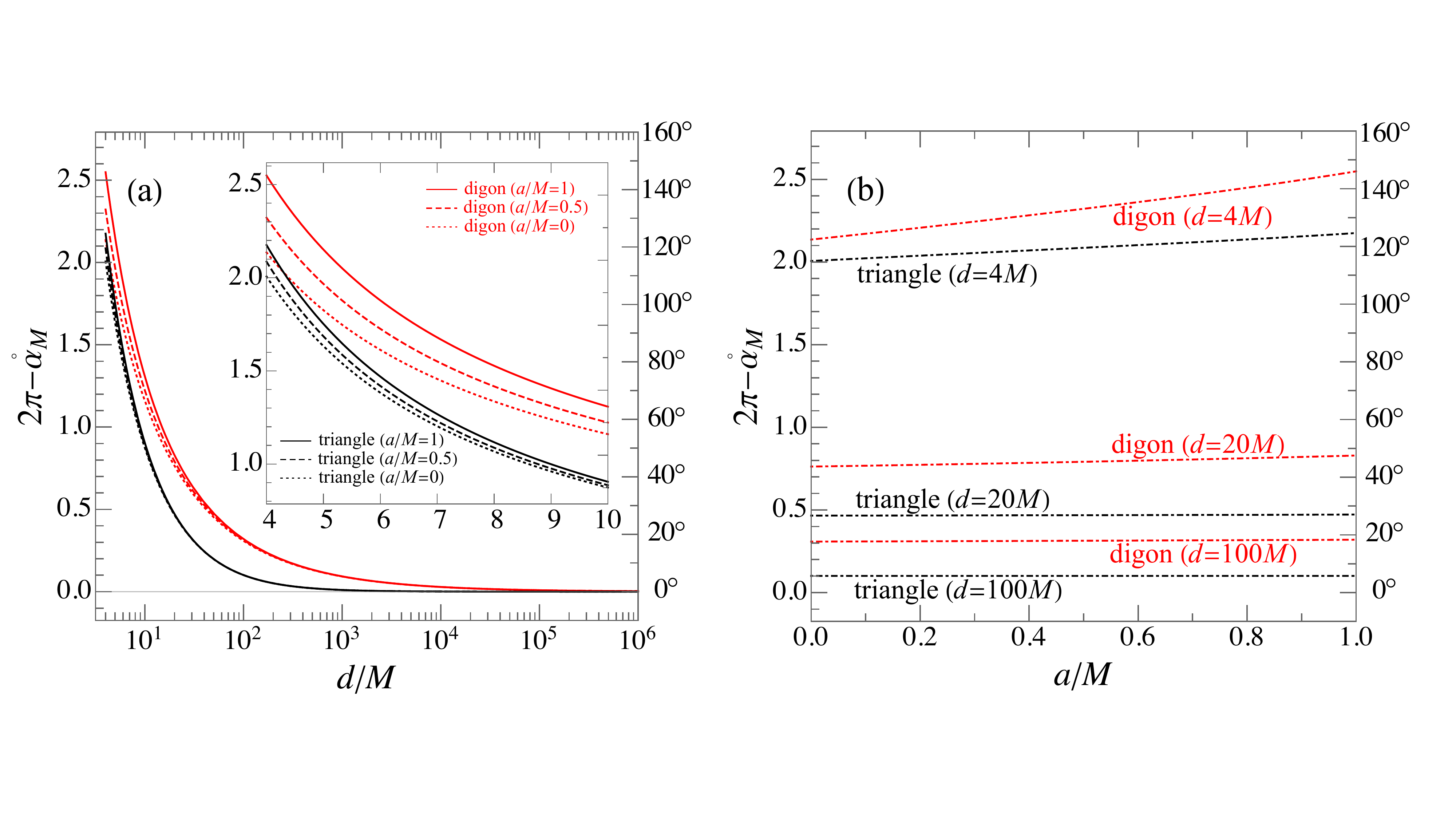}
\caption{
The excess of $2\pi$ over the Gaussian deflection angle, i.e., $2\pi\!-\!\mathring{\alpha}_{M}$, is shown as functions of distance $d$ and spin $a$ in panels (a) and (b), respectively. Here, $d$ represents the distance of each vertex of a regular geodesic polygon to its center, which is exactly the same as Figure~\ref{fig:triangle}. Results from the lensing patches in the shapes of a regular digon and triangle are colored red and black, respectively. 
}
\label{fig:GaM}
\end{figure}

Firstly, let us return to the regular digon; see the left panel of Figure~\ref{fig:digon} for a side-on view. Clearly, in this scenario, we have $D_{\rm LS}=D_{\rm L}=d$, and $D_{\rm S}=2d$.  Intrinsically, the Gaussian bending measure $\mathring{\alpha}_{M}$ is determined by the non-flatness of the physical surface. 
As a result, the excess of $2\pi$ over the Gaussian bending measure, i.e., $2\pi\!-\!\mathring{\alpha}_{M}$, can serve as a measurable quantity that describes how much the physical surface deviates from being flat. Panel (a) of Figure~\ref{fig:GaM} shows it as a function of the distance $d$ for $a/M=0,\,0.5,\,1$, respectively. This measurable quantity decreases significantly from $\sim100^{\circ}$ with increasing $d$, eventually approaching zero. Especially at $d\sim3.5\times10^{13}M$, it decreases to $\sim0.1^{\prime\prime}$, which is the level of accuracy at which Sir Eddington's expedition confirmed Einstein's prediction about the gravitational bending of light during a total eclipse of the Sun.

Additionally, the quantity $2\pi\!-\!\mathring{\alpha}_{M}$ increases with $a$, as illustrated in Figure~\ref{fig:GaM}. Especially in the close vicinity of a black hole, where $d/M\in[4,10]$, the disparity in $2\pi\!-\!\mathring{\alpha}_{M}$ between $a/M=0,\,0.5,\,1$ can be as large as several degrees, which is at least 5 orders of magnitude higher than the accuracy of $\sim0.1^{\prime\prime}$; see panel (a) for details. In order to observe this more clearly, in panel (b) of Figure~\ref{fig:GaM}, we also plot the quantity $2\pi\!-\!\mathring{\alpha}_{M}$ as a function of the spin $a$ for $d/M=4, 20, 100$, respectively. As this panel shows, the trend of $2\pi\!-\!\mathring{\alpha}_{M}$ increasing with $a$ becomes more significant as $d$ decreases. For instance, the increase in $2\pi\!-\!\mathring{\alpha}_{M}$ as $a/M$ increases from $0$ to $1$ is $0.66^{\circ}$ for $d/M\sim100$, whereas it becomes $3.83^{\circ}$ for $d/M\sim20$; see table~\ref{tab:value} for details. These strong-field effects are of great significance and help to determine the spin $a$ and mass $M$ of the black hole in a direct and highly precise way. 

For comparison, we also investigate the regular triangle; refer to Figure~\ref{fig:triangle} for a side-on view. As depicted in panel (a) of Figure~\ref{fig:GaM}, the quantity $2\pi\!-\!\mathring{\alpha}_{M}$ follows a similar trend with the distance $d$ as the regular digon. 
However, with a same value of $a$, this quantity tends to have a much lower value than the regular digon. 
More precisely, it decreases more rapidly with $d$. 
For example, it has already decreased to $\sim0.1^{\prime\prime}$ when $d\sim2\times10^{7}M$. 
Similar to the regular digon, the trend of the increase in $2\pi\!-\!\mathring{\alpha}_{M}$ with $a$ is also evidently strengthened with decreasing $d$. 
Specifically, the change in $2\pi\!-\!\mathring{\alpha}_{M}$ from $a/M=0$ to $a/M=1$ is $18.4^{\prime\prime}$ for $d/M\sim100$ and $0.36^{o}$ for $d/M\sim20$ (see table~\ref{tab:value}). Even though these strong-field effects are not as strong as those measured in a corresponding regular digon, they can still be probed with the same level of accuracy, about $\sim0.1^{\prime\prime}$, as measurements made over a century ago.

\begin{table}
\caption{Typical values of $2\pi\!-\!\mathring{\alpha}_{M}$ for experiments on the Gaussian bending of light in regular digons and triangles with a specified $d$. In the table, $\digon$ and $\Delta$ represent a regular digon and triangle, respectively. In each experiment, these values are shown in separate columns for $a/M=0$ and $a/M=1$, and they can be well approximated by a function when $d/M\gg10^4$, with the specific form of that function provided in the last row. 
}
\label{tab:value}
\vspace{2mm}
\centerline{
\begin{tabular}{|c|c|c|c|c|}
\hline
$2\pi\!-\!\mathring{\alpha}_{M}$&$a/M=0$ ($\digon$)~&~$a/M=1$~($\digon$)~&~$a/M=0$~($\Delta$)~&~$a/M=1$~($\Delta$)~\\
\hline
\!\!\!\!$d/M=4$&$122.38^{\circ}$&$146.04^{\circ}$&$114.96^{\circ}$&$124.64^{\circ}$\\
\hline
\!\!$d/M=20$&$43.68^{\circ}$&$47.51^{\circ}$&$26.69^{\circ}$&$27.05^{\circ}$\\
\hline
$d/M=10^2$&$17.67^{\circ}$&$18.33^{\circ}$&$5.79^{\circ}$&$5.80^{\circ}$\\
\hline
$d/M=10^3$&$5.27^{\circ}$&$5.33^{\circ}$&$35.62^{\prime}$&$35.62^{\prime}$\\
\hline
$d/M=10^4$&$1.635^{\circ}$&$1.641^{\circ}$&$3.571^{\prime}$&$3.572^{\prime}$\\
\hline
$d/M=10^5$&$30.84^{\prime}$&$30.87^{\prime}$&$21.43^{\prime\prime}$&$21.44^{\prime\prime}$\\
\hline
$d/M\gg10^5$&\!\!$162.1^\circ\left(d/M\right)^{-1/2}$&\!\!$162.1^\circ\left(d/M\right)^{-1/2}$\!\!&\!\!$595.3^\circ ({d/M})^{-1}$\!\!&\!\!$595.3^\circ ({d/M})^{-1}\!\!$\\
\hline
\end{tabular}
}
\end{table}

Note that for a given $d$, there are two distinct values of $2\pi\!-\!\mathring{\alpha}_{M}$ for regular {\it digons} and {\it triangles}; more details about the disparity can be found in figure~\ref{fig:GaM}. We summarise some typical values in table~\ref{tab:value}. For instance, at large distance $d$, the two different values can be approximated by the functions $162.1^\circ\left(d/M\right)^{-1/2}$ and $595.3^\circ ({d/M})^{-1}$, respectively, which are derived from fitting numerical values for $d/M\gg10^4$ with relative deviations of less than $\sim0.1\%$.
By comparing these two values, we can extract information about the basic parameters $\left(a, M\right)$ of the black hole and the distance $d$ to the black hole when performing two experiments in the circular orbit of radius $d$, based on the Gaussian bending measure of light in the digon and triangle, respectively. For a stellar black hole of $M\sim3 \,M_{\odot}$, 
these experiments can be done on the scale of $d\sim4-100 ~M\sim 15-450$ km.

As a vital and active area of research, the detection of PBHs could offer new insights into the nature of dark matter, the origin of the universe, cosmic evolution, and fundamental laws in gravitational physics. In detecting PBHs, one can design a meter-scale experiment.
For an Earth-mass PBH, with $M\sim10^{-6}~M_{\odot}$, the scale $d\sim10^{2}~M$ can be as small as a few meters, allowing for the measurement of the PBH's spin and mass with a triangle (or digon) of devices, to a lower accuracy of $\sim18.4^{\prime\prime}$ ($\sim0.66^\circ$). In contrast, for a comet-mass PBH, with $M\sim10^{-17}~M_{\odot}$ ($\sim10^{-11}~M_{\odot}$), the scale $d\sim10^{13}~M$ ($\sim10^{7}~M$) can be as large as a few meters, enabling the probe of the PBH with a digon (or triangle) of devices, to a higher accuracy of $\sim0.1^{\prime\prime}$.
The scale is surprisingly small, far more than expected. In the near future, we could design such experiments to search for the PBHs with $M\sim10^{-17}-10^{-6}~M_{\odot}$ in our solar system. Note that the masses involved are so large that Hawking radiation becomes negligible.
Conversely, through these experiments, we can also understand the gravity within our solar system, test various theories of gravity, or validate our theory by measurements of the gravitational field surrounding a mass, even if no PBHs are detected. 
Nonetheless, developing new means for the detection of such PBHs and the direct measurement of their mass and spin has been a longstanding challenge in astronomy. Our global theory of Gaussian bending has now laid a solid theoretical foundation for their development.

For actual measurements of the Gaussian bending measure, 
the geodesic polygon may deviate from a regular shape, despite its vertices being located in the same circular orbit. 
However, from the perspective of symmetry, we instinctively conjecture that {\it the value of $2\pi\!-\!\mathring{\alpha}_{M}$ will reach its minimum when the geodesic polygon is regular if all its vertices are located at the same distance $d$ to the center of the black hole}, which is already verified in a few instances by our numerical calculations. 
For instance, the regular digon can be likened to a special isosceles triangle, with two sides of equal length and the third side being infinitesimally small. 
For a given $d$, the value of $2\pi\!-\!\mathring{\alpha}_{M}$ is smaller for the regular triangle compared to the special isosceles triangle. 
How to prove this conjecture in general is actually a variational problem of the total Gaussian curvature $K_{\rm tot}$ as a functional of two functions with variable boundaries. To solidly prove it, we need to analytically solve the non-linear Euler equations involving at least two order derivatives. 
After substituting the Kerr metric components into these Euler equations, it can be found that it is almost impossible to prove it in an analytical way. 
Note that this is highly possible, although we are unable to prove it rigorously. 
Therefore, the strong-field effects of a black hole can be expected to be stronger than what we have calculated over regular polygons. 
In practice, experiments may be done by moving devices as local observers who are passing by the vertices. Thus, the values measured by the moving devices for the external or interior angles must be made relativistic corrections. In fact, the ultimately measured external or interior angle by these devices at each vertex can be corrected using general relativistic aberration relationships, accounting for corrections due to their positions and velocities~\cite{Lake2013,Zhang:2021ygh}. Additionally, we can install an atomic interferometer on each device to measure the gravitational field strength to a relative accuracy of $\sim10^{-10}$~\cite{Nan2018,Myszkowski:2022qho} and put extra constraints on positions and velocities. 
If these moving devices are subject to any extra forces, we need to make the geodesic correction $\Delta_{g}$ to the Gaussian bending measure $\mathring{\alpha}_{M}$. In any case, using the formula~\eqref{eq:GDA2}, we can determine the Gaussian bending measure from these external or interior angles.

At present, there are a variety of MG theories being suggested as solutions to some known shortcomings of GR. 
According to these theories, their physical differences from GR may be noticeable, particularly in how they behave in the strong-field limit. 
Subsequently, relevant studies have been conducted to delve into these differences through the gravitational bending of light \cite{Sanchez:2025dvb,Takizawa:2020dja,Sanchez:2023ckq,Jusufi:2017hed}.
For example, in the 4-dimensional spacetime theory of Einstein-Gauss-Bonnet gravity, 
there is a static, spherically symmetric metric solution that resembles the Schwarzschild metric at a small radius~\cite{Glavan20}. 
Nonetheless, this metric solution exhibits behavior more akin to the de Sitter metric in the large-radius region, 
resulting in a gravitational field that becomes very strong in the distant region, which is somewhat similar to that shown in the SdS spacetime~\cite{Zhang:2021ygh}\footnote{There is an ongoing discussion about the contribution of $\Lambda$ to the bending angle~\cite{RI2007,Ishak:2008ex,Arakida:2011ty}.}. 
Thus, it is not asymptotically flat. Accordingly, the usual deflection angle is clearly no longer measurable~\cite{RI2007,Ishak:2008ex,Arakida:2011ty}, as mentioned above. 
Consequently, the traditional theory of gravitational lensing is no longer applicable in the spacetime associated with this new solution, 
but our global theory remains valid. 
Recently, the accuracy of $\sim10^{-5}$ arcseconds has been achieved by GRAVITY~\cite{GRAVITY18}, 
which is often used to detect the events of gravitational microlensing. 
Thus, the current level of accuracy is at least 4 orders of magnitude higher than the level of $\sim0.1^{\prime\prime}$, 
greatly enhancing the potential for exploring physics beyond GR. 
To the current level of accuracy, 
we lack effective means to test these theories of gravity model-independently 
through measurements of gravitational bending conducted by local detectors in singular strong-field regions. 
More exactly, there was never a well-defined bending measure over singular spacetime regions, 
resulting in the lack of strict bending formulas for extracting information independently of coordinates from these local measurements in the highly-curved regions.
Now, we have been able to develop such means by enhancing the predicted new effects from MG theories through the proper selection of the outer boundary, as mentioned previously for the direct probe of dark energy, based on the global theory of Gaussian bending.

\section{Conclusions}
\label{sec:6}

In this work, we globally extended the Gaussian bending measure to singular spacetime regions, developed its global theory, 
and illustrated its applications in general stationary spacetimes.
(1). We defined the global Gaussian bending measure over a singular region without depending on the choice of coordinates and shapes of the region. 
Accordingly, we derived the global bending formula, along with a measurement formula, for this measure through validating the Gauss-bonnet theorem over the chosen region with multiple singularities. Specifically, we eliminated the singularities along the inner boundaries of the physical surface, forming various geometric ``holes'' in the chosen region, 
and then globally extended the theory of Gaussian bending to the region with singularities. In particular, we demonstrated that the spacetime region inside a black hole does not contribute to the global bending measure in the Kerr spacetime. 
(2). We investigated the global Gaussian bending measure of both massless and massive messengers in the most general stationary spacetime and explored its potential applications in understanding the astronomical phenomena, such as Einstein rings and the shadow imaging of black holes, as well as in designing future experiments that involve various polygonal lensing regions. Note that these experiments can be performed on very small scales in our Solar system, not restricted to the vicinity of a distant black hole. 
(3). We exemplified the global theory for the gravitational bending of light in the equatorial plane of a Kerr black hole, and investigated the strong-field effects of the black hole in its immediate vicinity. For instance, we depicted the Gaussian bending measure as a function of the spin and mass of the black hole. 
(4). We illustrated the potent strong-field effects on the Gaussian bending of light over geodesic polygons, such as the regular digons and triangles, 
enabling the extraction of precise information about the black hole. 
In the future,  the global theory could assist us in probing the unknown physics within the general theory of relativity and testing modified theories of gravity through local measurements in highly-curved spacetime regions.

\acknowledgments
We appreciate valuable comments and constructive suggestions from the anonymous reviewers that helped us improve the manuscript. This work is supported by the National Program on Key Research and Development Project from the Ministry of Science and Technology of China (Grant No. 2021YFA0718500), as well as the funding from the Chinese Academy of Sciences (Grant Nos. E329A3M1 and E3545KU2), the Institute of High Energy Physics (Grant No. E25155U1), and the National Natural Science Foundation of China (Grant No. 12235001).


\newpage
\appendix

\section{The global Gauss-Bonnet formula in singular lensing}
\label{app:T}
In this section, we will rederive the full version of the global Gauss-Bonnet formula from the local version of the Gauss-Bonnet theorem. Based on the detailed derivations, we will provide further insights into global Gaussian bending in singular spacetime regions.

Recall the local Gauss-Bonnet theorem. It establishes a connection between local and global properties of curves and surfaces,
which can be simply expressed as~\cite{Carmo16,Chern00}
\begin{eqnarray}
\label{eq:GBT0}
\begin{array}{rcl}
\displaystyle
\sum_{k}\int^{\lambda_{k+1}}_{\lambda_{k}}\kappa_{g}\mathrm{d} \lambda+\int\!\!\!\int_{\rm A}\,K\mathrm{d} \sigma+\sum_{k}\alpha_{k}=2\pi,
\hspace*{0mm}
\end{array}
\end{eqnarray}
where $\rm{A}$ is a simple singularity-free area with boundary $\gamma=\partial{\rm A}$, and $\kappa_{g}=\kappa_{g}\left(\lambda\right)$ is the geodesic curvature of the regular arcs of $\gamma$. Here, $\gamma$ is positively oriented, parametrised by arc length $\lambda$. For $\lambda=\lambda_{k}$, one has $\gamma_{k}=\gamma\left(\lambda_{k}\right)$, where  $\gamma_{k}$ is the $k-$th vertex of the boundary $\partial{\rm D}$.
If $\gamma$ is made up of geodesic segments, $\kappa_{g}\equiv0$.

Come back to Figure~\ref{fig:gdp1}. For $k=0, ... , \aleph-1$, let $\alpha_{k}$ ($\beta_{k}$) be the external (interior) angles of $\mathring{D}$ at its vertex $\gamma_{k}$, where $\aleph$ represents the total number of vertices.
For any $k^{*}\in\{i,j,l,m\}$, let $\tilde{\alpha}^{\rm o}_{k^{*}}$ ($\beta^{\rm o}_{k^{*}}$) and $\alpha^{\rm o}_{k^{*}}$ ($\tilde{\beta}^{\rm o}_{k^{*}}$) be the external (interior) angles of ${ D_{\rm u}}$ and ${ D_{\rm d}}$ at their respective vertices $\gamma_{k^{*}}$, respectively.
Then, according to the (local) Gauss-Bonnet theorem, one has
\begin{eqnarray}
\label{eq:GBT1}
\nonumber
\tilde{\alpha}^{\rm o}_{i}\!&\!+\!&\!\tilde{\alpha}^{\rm o}_{j}\!+\!\tilde{\alpha}^{\rm o}_{l}\!+\!\tilde{\alpha}^{\rm o}_{m}\!+\!\int_{C_{jm}}\kappa_{g}\,\mathrm{d} \lambda\!+\!\int_{C_{il}}\kappa_{g}\,\mathrm{d} \lambda\\
\!&\!+\!&\!\!\sum_{k^{\prime}}\alpha_{k^{\prime}}\!+\!\sum_{k^{\prime}}\int^{\lambda_{k^{\prime}+1}}_{\lambda_{k^{\prime}}}\kappa_{g}\,\mathrm{d} \lambda\!+\!\int\!\!\!\int_{D_{\rm u}}\,K\,\mathrm{d} \sigma=2\pi,~
\hspace*{0mm}
\end{eqnarray}
where $k^{\prime}\notin\{i,j,l,m\}$. 
Similarly, 
\begin{eqnarray}
\label{eq:GBT2}
\nonumber
\alpha^{\rm o}_{i}\!&\!+\!&\!\alpha^{\rm o}_{j}\!+\!\alpha^{\rm o}_{l}\!+\!\alpha^{\rm o}_{m}\!+\!\int_{C_{mj}}\kappa_{g}\,\mathrm{d} \lambda\!+\!\int_{C_{li}}\kappa_{g}\,\mathrm{d} \lambda\\
\!&\!+\!&\!\!\sum_{k^{\prime\prime}}\alpha_{k^{\prime\prime}}\!+\!\sum_{k^{\prime\prime}}\int^{\lambda_{k^{\prime\prime}+1}}_{\lambda_{k^{\prime\prime}}}\kappa_{g}\,\mathrm{d} \lambda\!+\!\int\!\!\!\int_{D_{\rm d}}\,K\,\mathrm{d} \sigma=2\pi,~
\hspace*{0mm}
\end{eqnarray}
where $k^{\prime\prime}\notin\{ i,j,l,m,k^{\prime}\}$.
In general, we find
\begin{eqnarray}
\label{eq:alphaalphao}
\alpha_{k^{*}}=\tilde{\alpha}^{\rm o}_{k^{*}}+\alpha^{\rm o}_{k^{*}}-\pi,
\hspace*{0mm}
\end{eqnarray}
for any $k^{*}\in\{i,j,l,m\}$. Therefore,
\begin{eqnarray}
\label{eq:GBT3}
\nonumber
(\alpha_{i}&+&\pi)\!+\!(\alpha_{j}+\pi)\!+\!(\alpha_{l}+\pi)\!+\!(\alpha_{m}+\pi)\\
&\!+\!&\sum_{\bar{k}^{*}}\alpha_{\bar{k}^{*}}\!+\!\sum_{k}\int^{\lambda_{k+1}}_{\lambda_{k}}\kappa_{g}\,\mathrm{d} \lambda\!+\!\int\!\!\!\int_{\mathring{D}}\,K\,\mathrm{d} \sigma=2\pi\times2,~~~
\hspace*{0mm}
\end{eqnarray}
where $\bar{k}^{*}\notin\{i,j,l,m\}$. Here, we have used
\begin{eqnarray}
\label{eq:GBapp}
\int_{C_{jm}}\kappa_{g}\,\mathrm{d} \lambda\!+\!\int_{C_{il}}\kappa_{g}\,\mathrm{d} \lambda\!+\!\int_{C_{mj}}\kappa_{g}\,\mathrm{d} \lambda\!+\!\int_{C_{li}}\kappa_{g}\,\mathrm{d} \lambda=0.
\hspace*{0mm}
\end{eqnarray}
Furthermore, equation~\eqref{eq:GBT3} can be rewritten as 
\begin{eqnarray}
\label{eq:GBT4}
\sum_{k}\alpha_{k}\!&\!+\!&\!\sum_{k}\int^{\lambda_{k+1}}_{\lambda_{k}}\kappa_{g}\,\mathrm{d} \lambda\!+\!\int\!\!\!\int_{\mathring{D}}\,K\,\mathrm{d} \sigma=0,
\hspace*{0mm}
\end{eqnarray}
where $k=0, ... , \aleph-1$. 
Note that the global Gauss-Bonnet theorem~\cite{Carmo16,Chern00} can be used to obtain this formula by setting the Euler characteristic number $\chi=\chi(\mathring{D})$ to zero.
However, this theorem cannot be applied directly to the region $D$ with a singularity without first removing the singular sub-region, $D_{\rm c}$, and creating a geometric ``hole'' in the lensing patch $\mathring{D}$. In addition, a mathematically meaningful theorem does not imply that it has physically reasonable interpretations.
Therefore, we have to re-derive the Gauss-Bonnet formula~\eqref{eq:GBT4} over the lensing patch $D$ with a singularity 
step by step to ensure that every step has a clear physical meaning.
Here are a number of things about this formula~\eqref{eq:GBT4} when applying it to real situations: 
(i). The line segments, such as $C_{il}$ and $C_{jm}$ in equation~\eqref{eq:GBapp}, connecting a vertex of $\partial{D}$ to one point on $\partial{D}_{\rm c}$ are not necessarily geodesic. (ii). $\partial{\mathring D}$ contains two boundaries, $\partial{D}_{\rm c}$ and $\partial{D}$, in topology. For simplicity, it can be formally denoted as $\partial{\mathring D}=\partial{D}-\partial{D}_{\rm c}$. Both of them should be made up of geodesic segments in a realistic physical situation. (iii). The surface integral of the Gaussian curvature must be calculated over the singularity-free region $\mathring{D}$ rather than the region $D$ with a singularity. 
(iv). When comparing the Gauss-Bonnet formulas~\eqref{eq:GBT0} and~\eqref{eq:GBT4}, a reduction of 2$\pi$ can be identified on the right-hand side, while the left-hand side remains unchanged in the form. In general, the creation of one more geometric ``hole'' in the region will lead to a further reduction of 2$\pi$ on the right-hand side of the Gauss-Bonnet formula.

Besides, there may be more than one singularity in the lensing patch $D$. 
Let $h$ represent the number of singularities. 
Using the same method as above, we can remove singular sub-regions $D^{r}_{\rm c}, r=0, 1,~...~,h-1$, around these singularities to create various shapes of geometric ``holes'', 
resulting in a singularity-free region, 
\begin{eqnarray}
\label{eq:mathringD2}
\begin{array}{rcl}
\displaystyle
\mathring{D}=D-D_{\rm c}=D-\bigcup\limits_{r\in\{0, ... , h-1\}} D^{r}_{\rm c},~~ 
\end{array}
\end{eqnarray}
which is now extended to incorporate more than one geometric ``hole''. 
Here, if $D$ has no singularities, $D_{\rm c}$ becomes empty. 
Currently, $h$ can be identified as the number of geometric ``holes'' in $\mathring{D}$. 
In geometry, the lensing patch $\mathring{D}$ is an orientable two-dimensional surface with multiple topological boundaries, 
including $\partial{D}^{r}_{\rm c}, r=0, 1,~...~,h-1$, and $\partial{D}$. 
Let $\partial{\mathring D}$ denote the boundary set of the region $\mathring{D}$, 
which can be formally represented as 
\begin{eqnarray}
\label{eq:mathringD2bound}
\begin{array}{rcl}
\displaystyle
\partial{\mathring D}=\partial{D}-\sum_{r}\partial{D}^{r}_{\rm c}, 
\end{array}
\end{eqnarray}
where $\partial{D}$ will be referred to as the outer boundary of the singularity-free region $\mathring{D}\subset\Sigma$, and $\partial{D}^{r}_{\rm c}$ as the $r$-th inner boundary hereafter. In topology, $\mathring{D}$ is homeomorphic to a 2-sphere with $g$ handles and $b=h+1$ ``holes'', where $g$ is the genus of $\mathring{D}$. 
Accordingly, the Euler characteristic number is $\chi=2-2 g-b$~\cite{Nakahara2003}. Therefore, we have $\chi=1-2 g-h$. 
After creating the region $\mathring{D}$ with $h$ ``holes'', we connect two points (or vertices) on some $\partial{D}^{r}_{\rm c}$ and two points (or vertices) of $\partial{D}$ with line segments, respectively, such that $\mathring{D}$ can be cut into two main parts: one with a single geometric ``hole'' and the other with $h-1$ ``holes''. Presume the validity of the global Gauss-Bonnet formula~\cite{Carmo16,Chern00} for the second part in advance. 
Then, by following the same procedure as described in equations~\eqref{eq:GBT1} to~\eqref{eq:GBT4}, we can validate the full version of the global Gauss-Bonnet formula by induction for the lensing patch $D$ with singularities, exactly as follows: 
\begin{eqnarray}
\label{eq:GBT5}
\sum_{k}\alpha_{k}\!+\!\sum_{k}\int^{\lambda_{k+1}}_{\lambda_{k}}\kappa_{g}\,\mathrm{d} \lambda\!+\!\!\int\!\!\!\int_{\mathring{D}}\,K\,\mathrm{d} \sigma=2\pi\,\chi(\mathring{D}),~~~~~~
\hspace*{0mm}
\end{eqnarray}
where the surface integral of the Gaussian curvature is performed over the singularity-free region $\mathring{D}$. 
In any case, adding an extra ``hole'' in the region $D$ leads to an additional reduction of $2\pi$ on the right-hand side, which is utilised in the process of induction. 
Note about the derivation of this global formula that all the steps remain valid even if the boundaries of the region $\mathring{D}$ are not composed of geodesic line segments. Based on this formula, we find a relationship between the global properties of the lensing patch $\mathring{D}$, 
including its total curvatures and the number $h$ of its geometric ``holes'', 
and the local properties of $\partial{\mathring D}$, such as its external (or interior) angles.

\section{A global theorem on Gaussian bending}
\label{app:A}
In this section, we address the specific case of a single singularity. 
In this case, we will prove that the Gaussian bending measure can be calculated as the sum of a topological invariant and a line integral along the outer boundary of a specific singular region, as shown in equation \eqref{GDAob}. In this way, the bending measure depends only on the outer boundary, unaffected by the inner boundary.

Assume the physical surface $\Sigma$ is orthogonally parametrised by the local coordinates $\left(u, \upsilon\right)$, associated with the following metric~\cite{Carmo16,Chern00}, 
\begin{eqnarray}
\label{eq:d2sigma_2d}
\begin{array}{rcl}
\displaystyle
\mathrm{d} \sigma^{2}&=&{E}\,\rm{d}{u}^{\rm{2}}+{G}\,\rm{d}{\upsilon}^{\rm{2}},
\end{array}
\end{eqnarray}
by which the Gaussian curvature $K$ can therefore be simply expressed as
\begin{eqnarray}
\label{eq:GaussKappB}
\begin{array}{rcl}
\displaystyle
K=-\frac{1}{\sqrt{EG}}\left(\left(\frac{(\sqrt{E})_{\upsilon}}{\sqrt{G}}\right)_{\!\upsilon} +\left(\frac{(\sqrt{G})_{u}}{\sqrt{E}}\right)_{\!u}\right), 
\end{array}
\end{eqnarray}
which only makes sense when $E>0$ and $G>0$. 
Now, let us consider the singularity-free region $\mathring{D}\subset\Sigma$ with one geometric ``hole'', such that $\chi(\mathring{D})=0$. 
It follows that the total curvature $K_{\rm{tot}}$ can be described in the following form, 
\begin{eqnarray}
\nonumber
K_{\rm{tot}}&=&-\oint_{\partial{\mathring{D}}}\left(-\frac{(\sqrt{E})_{\upsilon}}{\sqrt{G}}\,\rm{d}{u}+\frac{(\sqrt{G})_{u}}{\sqrt{E}}\,\rm{d}{\upsilon}\right)\\
\label{eq:totalappB}
\nonumber
&=&-\oint_{\partial{D}}\left(-\frac{(\sqrt{E})_{\upsilon}}{\sqrt{G}}\,\rm{d}{u}+\frac{(\sqrt{G})_{u}}{\sqrt{E}}\,\rm{d}{\upsilon}\right)\\
&&+\oint_{\partial{D}_c}\left(-\frac{(\sqrt{E})_{\upsilon}}{\sqrt{G}}\,\rm{d}{u}+\frac{(\sqrt{G})_{u}}{\sqrt{E}}\,\rm{d}{\upsilon}\right),~~~~~~~~
\hspace*{0mm}
\end{eqnarray}
where $\partial{D}_{c}$ is the inner boundary, while $\partial{D}$ is the outer boundary. 
Note that the integral is taken with a positive orientation in the counterclockwise direction. 
Mathematically, the inner and outer boundaries are actually two closed contours over which the two line integrals can be calculated, respectively. 
We denote the inner and outer contours as $L_{a}=\partial{D}_{c}$ and $L_{b}=\partial{D}$, respectively. 
Strictly speaking, these two contours are topologically equivalent, as they can be connected by a homeomorphism. Additionally, a different contour can always be obtained from one of them by a new homeomorphism. Similarly, many other contours can also be obtained through different homeomorphisms, finally forming a set of contours denoted as $\mathfrak{L}$. 
By definition, $L_{a}\in\mathfrak{L}$ and $L_{b}\in\mathfrak{L}$. 
For any $L \in \mathfrak{L}$, let us redefine its external angle $\breve{\alpha}_{k}$ at the vertex $\gamma_{k}$ to be positive in the counterclockwise direction. 
By this definition, $\breve{\alpha}_{k}=-\alpha_{k}$ at the vertex $\gamma_{k}$ of the inner contour $L_{a}$, while $\breve{\alpha}_{k}=\alpha_{k}$ at the vertex $\gamma_{k}$ of the outer contour $L_{b}$. 
In realistic physics, the two contours are made up of simple geodesic segments, respectively. Thus, $\kappa_{g}=0$. 
Subsequently, by equation~\eqref{eq:GDA2}, one has
\begin{eqnarray}
\sum\limits_{\gamma_{k}\in L}\alpha_k=\oint_{\partial{\mathring{D}}}\left(-\frac{(\sqrt{E})_{\upsilon}}{\sqrt{G}}\,\rm{d}{u}+\frac{(\sqrt{G})_{u}}{\sqrt{E}}\,\rm{d}{\upsilon}\right)\,, 
\end{eqnarray}
which is valid if and only if the Gauss-Bonnet theorem holds true. 
Then, substituting equation~\eqref{eq:totalappB} into this equation yields
\begin{eqnarray}
\sum\limits_{\gamma_{k}\in L_a}\breve{\alpha}_{k}-\sum\limits_{\gamma_{k}\in L_b}\breve{\alpha}_{k}=&\oint_{L_a}\left(-\frac{(\sqrt{E})_{\upsilon}}{\sqrt{G}}\,\rm{d}{u}+\frac{(\sqrt{G})_{u}}{\sqrt{E}}\,\rm{d}{\upsilon}\right)\nonumber\\
&-\oint_{L_b}\left(-\frac{(\sqrt{E})_{\upsilon}}{\sqrt{G}}\,\rm{d}{u}+\frac{(\sqrt{G})_{u}}{\sqrt{E}}\,\rm{d}{\upsilon}\right)\,. 
\end{eqnarray}
Now we can introduce a mapping $f:\mathfrak{L}\rightarrow \mathbb{R}$, defined as follows: 
\begin{eqnarray}
f(L_a)\equiv \sum\limits_{\gamma_{k}\in L_a}\breve{\alpha}_{k}-\oint_{L_a}\left(-\frac{(\sqrt{E})_{\upsilon}}{\sqrt{G}}\,\rm{d}{u}+\frac{(\sqrt{G})_{u}}{\sqrt{E}}\,\rm{d}{\upsilon}\right). 
\end{eqnarray}
Therefore, we have
\begin{eqnarray}
\label{eq:fafb}
f(L_a)=f(L_b)\,. 
\end{eqnarray}
In fact, both $L_a$ and $L_b$ can be chosen at our convenience. 
For any two elements, $L_a, L_b\in\mathfrak{L}$, the formula~\eqref{eq:fafb} remains valid. 
Hence, $f$ is a mapping of the set $\mathfrak{L}$ onto a constant. As a result, we obtain
\begin{eqnarray}
\label{fL.const}
f(L)\equiv \sum\limits_{\gamma_{k}\in L}\breve{\alpha}_{k}-\oint_{L}\left(-\frac{(\sqrt{E})_{\upsilon}}{\sqrt{G}}\,\rm{d}{u}+\frac{(\sqrt{G})_{u}}{\sqrt{E}}\,\rm{d}{\upsilon}\right)=\mathfrak{Z}_{0}, 
\end{eqnarray}
for any $L \in \mathfrak{L}$. It is important to emphasize that the constant $\mathfrak{Z}_{0}$ is a topological invariant, and its exact value may vary between different physical surfaces. 
Then, according to equation~\eqref{eq:GDA2-4}, we find
\begin{eqnarray}
\label{fL.const2}
\mathring{\alpha}_{M}=\oint_{L}\left(-\frac{(\sqrt{E})_{\upsilon}}{\sqrt{G}}\,\rm{d}{u}+\frac{(\sqrt{G})_{u}}{\sqrt{E}}\,\rm{d}{\upsilon}\right)+\mathfrak{Z}_{0}, 
\end{eqnarray}
where $L$ is an arbitrary element in the set $\mathfrak{L}$. Note that this is true only if the Gauss-Bonnet theorem is valid. 
In particular, it still holds well in the case of the outer boundary, i.e., $L=L_{b}$. This clearly means that $\mathring{\alpha}_{M}$ is independent of the inner boundary $L_{a}$. 
The value of $\mathfrak{Z}_{0}$ can be determined through calculations over a specific contour. 
For instance, in the equatorial plane~\eqref{eq:EFG} of a Kerr black hole, we can choose a contour with its vertices located at spatial infinity. 
This contour can be a geodesic polygon. Clearly, the sum of the external angles of this geodesic polygon is $2\pi$, as the Kerr spacetime is asymptotically flat. 
Over this geodesic polygon, the line integral in equation~\eqref{fL.const} also equals $2\pi$, because $E\to1$ and $G\to r^{2}$ with $r\to+\infty$. 
Thus, we always have $\mathfrak{Z}_{0}=0$ in the equatorial plane of a Kerr black hole, which is also confirmed by the results presented in Figure~\ref{fig:GaM}.



\end{document}